\documentstyle[12pt,epsfig,amssymb]{article}
\begin{document}
\setlength{\parskip}{0.45cm}
\setlength{\baselineskip}{0.75cm}
\begin{titlepage}
\begin{flushright}
CERN-TH/97-286 \\ 
DTP/97/106 \\
November 1997\\ 
\end{flushright}
\vspace{0.6cm}
\begin{center}
\Large
{\bf QCD Analysis of Unpolarized and Polarized} 

\vspace{0.1cm}
{\bf {$\boldmath{\Lambda}$} - Baryon Production in} 

\vspace{0.1cm}
{\bf Leading and Next-to-Leading Order} 

\vspace{1.2cm}
\large
D.\ de Florian$^a$, M.\ Stratmann$^b$, W.\ Vogelsang$^a$\\

\vspace*{1.5cm}
\normalsize
{\it $^a$Theoretical Physics Division, CERN, CH-1211 Geneva 23, Switzerland}

\vspace*{0.1cm}
{\it $^b$Department of Physics, University of Durham, Durham DH1 3LE, England}

\vspace{1.5cm}
%
%%%%%%%%%%%%%%%%%
\large
{\bf Abstract} \\
\end{center}
\vspace{0.5cm}
\normalsize
We analyze experimental data for the production of $\Lambda$ baryons in 
$e^+e^-$ annihilation in terms of scale dependent, QCD evolved, $\Lambda$ 
fragmentation functions. Apart from the vast majority of the data for which
the polarization of an observed $\Lambda$ was not determined, we also 
consider the recent LEP measurements of the longitudinal 
polarization of $\Lambda$'s produced on the $Z$-resonance. 
Such data correspond to spin-dependent
fragmentation functions for the $\Lambda$. We point out that the present 
data are insufficient to satisfactorily fix these. We therefore suggest 
several different sets of fragmentation
functions, all compatible with present data, and study the prospects 
for conceivable future semi-inclusive deep-inelastic scattering
experiments to discriminate between them. We provide the complete 
next-to-leading order QCD framework for all the processes we consider.
\end{titlepage}
\newpage
%%%%%%%%%%%%%%%%%%%%%%%
\section{Introduction}
%%%%%%%%%%%%%%%%%%%%%%%
\noindent
Measurements of rates for single-inclusive $e^+e^-$ annihilation (SIA) into 
a specific hadron $H$,
\begin{equation}
e^+e^- \rightarrow (\gamma,\,Z) \rightarrow H\;X\;\;\;,
\end{equation}
play a similarly fundamental role as those of the corresponding crossed 
``space-like'' deep-inelastic scattering (DIS) process $ep\rightarrow e' X$. 
Their interpretation in terms of scale-dependent fragmentation functions 
$D_f^{H}(z,Q^2)$, the ``time-like'' counterparts of the parton distribution 
functions $f_H(x,Q^2)$ of a hadron $H$, provides a further important, 
complementary test of perturbative QCD. In analogy with the ``space-like'' 
case, $D_f^{H}(z,Q^2)$ is the probability at a mass scale $Q$ for finding 
a hadron $H$ carrying a fraction $z$ of the parent parton's momentum.
QCD completely predicts the $Q^2$-dependence of the process-independent
fragmentation functions $D_f^{H}(z,Q^2)$ via the Altarelli-Parisi evolution
equations, once a suitable non-perturbative hadronic input at some initial 
reference scale $\mu$ has been determined from the data. So far, only the 
fragmentation into the most copiously produced light mesons ($\pi,\,K$) has 
been the issue of a thorough QCD analysis \cite{bkk}. 

It is one of the purposes of this work to study along similar lines as in 
\cite{bkk} whether such a formalism also applies to the production of 
$\Lambda$-baryons. $\Lambda$'s are also produced at fairly large rates and, 
as for pions and kaons, the $Q^2$ range covered by present SIA experiments
\cite{newdata,slddata} allows a detailed quantitative QCD analysis. Recently 
measured $\Lambda$ production rates in semi-inclusive DIS (SIDIS) 
\cite{h1dis} provide an important testing ground for the fragmentation 
functions extracted from SIA data.

The production of $\Lambda$ baryons appears to be particularly  
interesting also from a different point of view. Contrary to spinless mesons 
like pions and kaons, the $\Lambda$ baryon offers the rather unique 
possibility to study for the first time spin transfer reactions.
The self-analyzing properties of its dominant weak decay 
$\Lambda \rightarrow p \pi^-$ and the particularly large asymmetry of the 
angular distribution of the decay proton in the $\Lambda$ rest-frame 
\cite{perkins} allow an experimental reconstruction of the $\Lambda$ spin. 
Over the past years ``spin physics'' has attracted an ever growing interest, 
as experimental findings \cite{mallot} have not always matched with 
``naive'' theoretical expectations, the Gourdin-Ellis-Jaffe sum rule 
\cite{gej} being the most prominent example here. Studies of $\Lambda$ 
polarization could provide a completely new insight into the field of spin 
physics whose theoretical understanding is still far from being complete 
despite recent progress, and they might also yield further information 
on the hadronization mechanism.

In \cite{burkjaf} a strategy was proposed for extracting in SIA 
the functions $\Delta D_f^{\Lambda}(z,Q^2)$ describing the fragmentation of 
a longitudinally polarized parton into a longitudinally polarized 
$\Lambda$ \cite{ji},
\begin{equation}
\label{eq:deltad}
\Delta D_f^{\Lambda}(z,Q^2) \equiv D_{f(+)}^{\Lambda (+)}(z,Q^2) -
 D_{f(+)}^{\Lambda (-)}(z,Q^2)\;\;\;,
\end{equation}
where $D_{f(+)}^{\Lambda (+)}(z,Q^2)$ $(D_{f(+)}^{\Lambda (-)}(z,Q^2))$ is 
the probability for finding a $\Lambda$ with positive (negative) helicity 
in a parton $f$ with positive helicity (by taking the sum instead of the
difference in (\ref{eq:deltad}) one recovers the unpolarized 
fragmentation function $D_f^{\Lambda}$). If the energy is far below 
the $Z$-resonance, one longitudinally polarized beam is required in order to
create a non-vanishing net polarization of the outgoing (anti)quark that 
fragments into the $\Lambda$, and to obtain a non-zero twist-two spin 
asymmetry. At higher energies, such as at LEP, even {\em{no}} beam 
polarization is required since the parity-violating $q\bar{q}Z$ coupling 
automatically generates a net polarization of the quarks. Here, ALEPH 
\cite{aleph}, DELPHI \cite{delphi}, and OPAL \cite{opal} have recently 
reported first results for the polarization of $\Lambda$'s produced on 
the $Z$-resonance.

Realistic models for the $\Delta D_{f}^{\Lambda}(z,Q^2)$ are also of 
particular relevance for reliable estimates of production rates and spin 
transfer asymmetries at present and future dedicated spin experiments. Here
the $\Delta D_{f}^{\Lambda}(z,Q^2)$ can be probed in SIDIS or photoproduction
in the current fragmentation region, $lp \rightarrow l' \Lambda X$, where
either a longitudinally polarized lepton beam or a polarized nucleon target
would be required. Such measurements can be carried out at HERMES 
\cite{hermes2} and are planned by the COMPASS \cite{compass} collaboration.
After the scheduled upgrade of the HERA electron ring with 
spin rotators in front of the H1 and ZEUS experiments, longitudinally
polarized electrons will be also available for high-energy $ep$ collisions, 
and similar measurements with polarized $\Lambda$'s in the final state 
could be performed here.
Furthermore, having also a polarized {\em{proton}} 
beam available at HERA \cite{polhera}
would allow the measurement of various different twist-2 asymmetries,
depending on whether the $e$ and/or the $p$ beam and/or the
$\Lambda$ are polarized, i.e., $\vec{e}p\rightarrow \vec{\Lambda} X$,
$e \vec{p}\rightarrow \vec{\Lambda} X$, and $\vec{e}\vec{p}\rightarrow 
\Lambda X$ (as usual, an arrow denotes a polarized particle). 

So far estimates for future $\Lambda$ experiments have relied on simple 
models \cite{nzar} or on Monte-Carlo simulations  
tuned with several parameters and parametrizations of 
scale-{\em{in}}dependent spin-transfer coefficients $C_f^{\Lambda}$ which 
link longitudinally polarized and unpolarized fragmentation functions 
via \cite{bravar}
\begin{equation}
\label{eq:ctrans}
\Delta D_f^{\Lambda}(z) = C_f^{\Lambda}(z) D_f^{\Lambda}(z)\;\;\;.
\end{equation}
Different phenomenological models for the $C_f^{\Lambda}$ exist. A first
one is based on the naive non-relativistic quark model where 
only $s$-quarks can contribute to the fragmentation processes that 
eventually yield a polarized $\Lambda$. Another approach goes back to  
estimates by Burkardt and Jaffe \cite{burkjaf,jaffe2} for a 
fictitious DIS structure function $g_1^{\Lambda}$ of the
$\Lambda$, for which sizeable negative contributions from $u$ and
$d$ quarks are predicted by analogy with the breaking of the 
Gourdin-Ellis-Jaffe sum rule \cite{gej} for the proton's $g_1^p$.
It is then assumed that such features also carry over to the 
``time-like'' case \cite{jaffe2} (see also \cite{bravar}). Of course, 
relations like (\ref{eq:ctrans}) cannot 
in general hold true in QCD. Due to the different $Q^2$-evolutions 
of $\Delta D_f^{\Lambda}$ and $D_f^{\Lambda}$, it cannot be correct
to assume scale independence of the $C_f^{\Lambda}$ in (\ref{eq:ctrans}), 
and therefore one has to specify a scale at which one implements such 
an ansatz.

It is the main purpose of this paper to address the 
issue of fragmentation into polarized $\Lambda$'s in a detailed QCD analysis. 
Here it will be possible for us to work even at next-to-leading order 
(NLO) accuracy, as the required spin-dependent ``time-like'' two-loop 
evolution kernels were derived recently \cite{tlpol}.
For the first time, we will provide some realistic sets of unpolarized and
polarized fragmentation functions for $\Lambda$ baryons.
A useful restrictive constraint when constructing
models for the $\Delta D_f^{\Lambda}$ is provided by the positivity 
condition (similarly to the ``space-like'' case), i.e.
\begin{equation}
\label{eq:positivity}
\left|\Delta D_f^{\Lambda}(z,Q^2)\right| \le  D_f^{\Lambda}(z,Q^2)\;\;,
\end{equation}
with the $D_f^{\Lambda}(z,Q^2)$ taken from the unpolarized analysis.
As the available sparse data from LEP are by far not sufficient to 
completely fix the $\Delta D_f^{\Lambda}(z,Q^2)$, we will propose 
several different sets of polarized fragmentation functions, all 
compatible with the LEP data. Some of the sets will be based on the ideas 
outlined in the previous paragraph. Our various proposed 
$\Delta D_f^{\Lambda}$ are particularly suited for estimating the 
physics potential of future experiments to determine the polarized 
fragmentation functions more precisely. We hence present detailed 
predictions for future SIDIS measurements at HERMES and the HERA collider.
In this context we also provide the necessary framework to calculate 
helicity transfer cross sections in SIDIS at NLO.

The remainder of the paper is organized as follows: in the next Section 
we develop the formalism for unpolarized SIA and discuss in detail
our analysis of leading order (LO) and NLO $\Lambda$ fragmentation functions.
In Section 3 we turn to the case of longitudinally polarized 
$\Lambda$ production and present our different 
conceivable scenarios for the $\Delta D_f^{\Lambda}(z,Q^2)$.
In Section 4 we compare our unpolarized distributions
with recent $\Lambda$ production data in SIDIS and study the potential of
present (HERMES) and future (HERA) spin physics experiments 
to discriminate between the different proposed sets of polarized 
fragmentation functions. Finally our results are summarized in Section 5. 
The Appendices collect the required unpolarized and polarized
NLO coefficient functions for SIA and SIDIS.
%
%%%%%%%%%%%%%%%%%%%%%%%%%%%%%%%%%%%%%%%%%%%%%%%%%%%%%%%
\section{Unpolarized $\Lambda$ Fragmentation Functions}
%%%%%%%%%%%%%%%%%%%%%%%%%%%%%%%%%%%%%%%%%%%%%%%%%%%%%%%
\noindent
In the last few years several experiments \cite{newdata,slddata} have reported 
measurements of the unpolarized cross section for the production of 
$\Lambda$ baryons, which allows an extraction of the unpolarized $\Lambda$ 
fragmentation functions required for constructing the polarization 
asymmetries and as reference distributions in the positivity constraint 
(\ref{eq:positivity}). We emphasize at this point that the wide range 
of c.m.s.\ energies covered by the data \cite{newdata,slddata} 
($14 \leq \sqrt{s} \leq 91.2$ GeV) makes a detailed QCD analysis that 
includes the $Q^2$-evolution of the fragmentation functions mandatory.

The cross section for the inclusive production of a hadron $H$ with 
energy $E_H$ in SIA at a c.m.s.\ energy $\sqrt{s}$, integrated over the 
production angle, can be written in the following way \cite{altarelli,fupe}:
\begin{equation}
\label{eq:cross}
\frac{1}{\sigma_{tot}} \frac{d\sigma^H}{dx_E} = \frac{1}{\sum_q \hat{e}_q^2}
\left[ 2\, F_1^{H}(x_E,Q^2) +  F_L^{H}(x_E,Q^2) \right] \; ,
\end{equation}
where $x_E=2 p_H\cdot q/Q^2 = 2 E_H/\sqrt{s}$ ($q$ being the 
momentum of the intermediate $\gamma$ or $Z$ boson, $q^2=Q^2=s$) and
\begin{equation}
\label{eq:sigmatot}
\sigma_{tot}=\sum_q \hat{e}_q^2\; \frac{4 \pi \alpha^2(Q^2)}{s} 
\left[1+\frac{\alpha_s(Q^2)}{\pi}\right]
\end{equation}
is the total cross section for $e^+e^- \rightarrow hadrons$ including its NLO
${\cal{O}}(\alpha_s)$ correction. The sums in (\ref{eq:cross}),
(\ref{eq:sigmatot}) run over the $n_f$ active quark flavours $q$, and 
the $\hat{e}_q$ are the corresponding appropriate electroweak charges
(see Appendix A for details).

To NLO accuracy the unpolarized ``time-like'' structure functions 
$F_1^{H}$ and $F_L^{H}$ in (\ref{eq:cross}) are given by
\begin{eqnarray}
\label{eq:f1nlo}
2 \, F_1^{H}(x_E,Q^2) &=& \sum_q \hat{e}_q^2\; 
\Bigg\{ \left[ D_q^H (x_E,Q^2) + D_{\bar q}^H (x_E,Q^2) \right] \\
&& \left. + \frac{\alpha_s(Q^2)}{2\pi} \left[ C_q^1 \otimes 
(D_q^H+D_{\bar q}^H) + C_g^1 \otimes D_g^H \right] (x_E,Q^2)
\right\} \; \; , \nonumber\\
\label{eq:flnlo}
F_L^{H}(x_E,Q^2) &=& \frac{\alpha_s(Q^2)}{2\pi} \sum_q \hat{e}_q^2 \;
\left[ C_q^L \otimes (D_q^H+D_{\bar q}^H) + C_g^L \otimes D_g^H 
\right](x_E,Q^2) \;\;,
\end{eqnarray}
with the convolutions $\otimes$ being defined as usual by
\begin{equation} \label{eq:convolut}
(C\otimes D)(x_E,Q^2) = \int^1_{x_E} \frac{dy}{y}\, 
C\left(\frac{x_E}{y}\right) D(y,Q^2)\;\;.
\end{equation}
The relevant NLO coefficient functions $C_{q,g}^{1,L}$ \cite{altarelli,fupe}
can also be found in Appendix A.

To determine the $Q^2$-evolution of the $D_f^H$ in 
Eqs.\ (\ref{eq:f1nlo}), (\ref{eq:flnlo}) it is as usual convenient to
decompose them into flavor singlet and non-singlet pieces by introducing the
densities $D_{q,\pm}^H$ and the vector
\begin{equation} \label{eq:startevol}
\vec{D}^H \equiv \left( \begin{array}{c} D_{\Sigma}^H \\ D_g^H \\ 
\end{array}\right) \; ,
\end{equation}
where
\begin{equation} 
\label{eq:nsdef}
D_{q,\pm}^H \equiv D_q^H \pm D_{\bar{q}}^H\;\;\; , \;\;\;
D_{\Sigma}^H \equiv \sum_q (D_q^H+ D_{\bar{q}}^H) \; .
\end{equation}
One then has the following non-singlet
evolution equations ($q,\tilde{q}$ being two different flavors):
\begin{eqnarray} 
\label{eq:ns1evol}
\frac{d}{d\ln Q^2} (D_{q,+}^H - D_{\tilde{q},+}^H) (z,Q^2) &=&
\left[ P_{qq,+}^{(T)} \otimes (D_{q,+}^H - D_{\tilde{q},+}^H)\right] 
(z,Q^2) \; , \\
\label{ns2evol}
\frac{d}{d\ln Q^2} D_{q,-}^H (z,Q^2) &=&
\left[P_{qq,-}^{(T)} \otimes D_{q,-}^H\right] (z,Q^2) \;\;.
\end{eqnarray}
The two evolution kernels $P_{qq,\pm}^{(T)}(z,\alpha_s(Q^2))$
become different beyond LO as a result of the presence of transitions 
between quarks and antiquarks. The singlet evolution equation reads
\begin{eqnarray} 
\label{eq:sing}
\frac{d}{d\ln Q^2} \vec{D}^H (z,Q^2) = \left[
\hat{P}^{(T)}\otimes \vec{D}^H\right](z,Q^2) \;\;\;,
\end{eqnarray}
where we write the singlet evolution matrix $\hat{P}^{(T)}$ as:
\begin{eqnarray}
\label{eq:pmatrix}
\renewcommand{\arraystretch}{1.3}
\hat{P}^{(T)} \equiv \left( \begin{array}{cc}
P_{qq}^{(T)} &  2n_f P_{gq}^{(T)} \\
\frac{1}{2n_f} P_{qg}^{(T)} & P_{gg}^{(T)} \\
\end{array}\right) \; .
\end{eqnarray}
To NLO, all splitting functions \cite{split,grvgam} 
in (\ref{eq:ns1evol})-(\ref{eq:pmatrix}) have the perturbative expansion
\begin{equation} 
\label{eq:expan}
P_{ij}^{(T)} (z,\alpha_s) = \left( \frac{\alpha_s}{2\pi} \right)
P_{ij}^{(T),(0)} (z) + \left( \frac{\alpha_s}{2\pi} \right)^2
P_{ij}^{(T),(1)} (z) \; .
\end{equation}
It should be noted that the evolution equations can be straightforwardly 
solved analytically in Mellin-$n$ space along the lines as, e.g.,
described in \cite{fupe}. The desired $D_f^H(z,Q^2)$ are then obtained
by a standard numerical Mellin-inversion. Needless to mention that
the corresponding LO expressions are entailed in 
Eqs.\ (\ref{eq:sigmatot})-(\ref{eq:flnlo}) by simply dropping all 
${\cal{O}}(\alpha_s)$ contributions and by evolving the $D_f^H(z,Q^2)$ in LO. 

In our numerical analysis we cannot include data with $x_E<0.1$, for two 
reasons: first of all, for small $x_E$, finite-mass corrections to 
Eq.\ (\ref{eq:cross}) proportional to $4 M_\Lambda^2/sx_E^2$ become more and 
more important, but are not accounted for in the calculation. There is also a 
more severe limitation set by the evolution equations outlined
above: the NLO ``time-like'' splitting functions $P_{gq}^{(T),(1)}(z)$ and 
$P_{gg}^{(T),(1)}(z)$ in (\ref{eq:pmatrix}) turn out be much more singular 
than their corresponding ``space-like'' counterparts as $z\rightarrow 0$.
While the leading small-$x$ terms in the ``space-like'' case are proportional
to $1/x$,
\begin{eqnarray}
\lim_{x\to 0} P^{(S)}_{gq} (x) &=& 
\frac{\alpha_s}{2\pi}\left(\frac{2C_F}{x}+\frac{\alpha_s}{2\pi} 
\frac{9C_F C_A-40 C_F T_f}{9x}\right)\,,\,\\
\lim_{x\to 0} P^{(S)}_{gg} (x) &=& 
\frac{\alpha_s}{2\pi}\left(\frac{2C_A}{x}+\frac{\alpha_s}{2\pi} 
\frac{12 C_F T_f-46 C_A T_f}{9x}\right)\,,
\end{eqnarray}
the ``time-like'' splitting functions show an even stronger negative behaviour 
for $z\rightarrow 0$ due to the dominant large logarithmic piece 
$\simeq \ln^2 z/z$ in the NLO part,
\begin{eqnarray}
\lim_{z\to 0} P^{(T)}_{gq} (z) &=&
\frac{\alpha_s}{2\pi}\left(\frac{2C_F}{z} - \frac{\alpha_s}{2\pi}  
\frac{4 C_F C_A}{z} \ln^2 z\right) \,,\,\\
\lim_{z\to 0} P^{(T)}_{gg} (z) &=&
\frac{\alpha_s}{2\pi}\left(\frac{2C_A}{z} - \frac{\alpha_s}{2\pi}  
\frac{4 C_A^2}{z} \ln^2 z\right) 
\end{eqnarray}
(see for example \cite{ellisbook}),
with the usual QCD colour factors $C_A=3$, $C_F=4/3$, and $T_f=T_R n_f= n_f/2$.
This singular behaviour of the ``time-like'' splitting functions is even so
strong that it may ultimately lead to {\em{negative}} 
NLO fragmentation functions in the course of the $Q^2$-evolution 
and hence to unacceptable 
{\it negative} cross sections at some value $x_E\ll 1$, even
if the evolution starts with positive distributions at the initial scale. 
Clearly, the description of fragmentation processes by perturbative 
QCD without resummation of small-$z$ logarithms breaks down for values 
of $x_E$ where this happens, and in order to avoid these severe problems 
we include as usual (see, e.g., \cite{bkk}) only data with $x_E>0.1$ in our 
analysis\footnote{We do not include data either that have been averaged 
experimentally over a large bin of $x_E$.}. 
 
As pointed out in \cite{burkjaf}, the QCD formalism is strictly speaking
only applicable to strongly produced $\Lambda$'s. A certain fraction of the
data~\cite{newdata,slddata} will, however, consist of 
secondary $\Lambda$'s resulting 
from $e^+ e^- \rightarrow \Sigma^0 X$ with the subsequent decay 
$\Sigma^0 \rightarrow \Lambda \gamma$, not to be included in the 
fragmentation functions \cite{burkjaf}. For simplicity, we will ignore 
this problem and (successfully) attempt to describe the full data samples by 
fragmentation functions that are evolved according to the QCD 
$Q^2$-evolution equations.

Unless stated otherwise, we will refer to both $\Lambda^0$ and 
$\bar{\Lambda}^0$, which are not usually distinguished in present $e^+e^-$ 
experiments \cite{newdata,slddata}, as simply ``$\Lambda$''. 
As a result, the obtained fragmentation functions always correspond to the sum
\begin{equation}
\label{eq:lamlambar}
D_f^\Lambda(x_E,Q^2) \equiv D_f^{\Lambda^0}(x_E,Q^2)+ 
D_f^{\bar{\Lambda}^0}(x_E,Q^2)\;\;\;.
\end{equation}
This also considerably simplifies the analysis, since no distinction between 
``favoured''  and ``unfavoured'' distributions is required. Since no precise 
SIDIS data are available yet, it is not possible to obtain individual 
distributions for all the light flavours separately, and hence some sensible 
assumptions concerning them have to be made. Employing naive quark model 
$SU_f(3)$ arguments and neglecting any mass differences between the  
$u$, $d$, and $s$ quarks, 
we {\it assume}\footnote{Fits allowed to be more general 
do not seem to improve the final $\chi^2/d.o.f.$} 
that all the light flavours fragment equally into $\Lambda$, i.e.
\begin{eqnarray}
\label{eq:ansatz}
D^{\Lambda}_u=D^{\Lambda}_d=D^{\Lambda}_s=D^{\Lambda}_{\bar{u}}=
D^{\Lambda}_{\bar{d}}=D^{\Lambda}_{\bar{s}} 
\equiv D^{\Lambda}_q\;\;.
\end{eqnarray} 
Needless to say that the $q$ and $\bar{q}$ fragmentation functions 
in (\ref{eq:ansatz}) are equal due to Eq.\ (\ref{eq:lamlambar}).  

At variance with the usual DIS case, {\it dis}continuous heavy quark
(HQ) fragmentation functions should be included at each heavy flavour 
threshold (see also \cite{bkk}). Anyway, the inclusion of the HQ contributions
essentially only leads to a change in the normalization of the light 
quark densities, which would just be larger if the HQ ones 
were not present. In our analysis we start the evolution of the HQ 
contributions at the mass of the corresponding HQ, but the precise value 
for this is anyhow irrelevant since all the data are in a region of 
$s>m_h^2$ ($h=c,b$).

For our analysis, we choose to work in the framework of the ``radiative 
parton model'' which is characterized by a rather low starting scale 
$\mu$ for the 
$Q^2$-evolutions. The ``radiative parton model'' has proven phenomenologically 
successful in the ``space-like'' case for both unpolarized \cite{grv} and 
polarized \cite{grsv} parton densities, and also in the ``time-like'' 
situation for photon fragmentation functions \cite{grvgam,opalg}.
 
At the initial scale ($\mu^2_{LO}=0.23\,{\mathrm{GeV}}^2$, 
$\mu^2_{NLO}=0.34\,{\mathrm{GeV}}^2$) we choose the following simple ansatz:
\begin{equation}
\label{eq:input}
D^{\Lambda}_f (z,\mu^2) = N_f\, z^{\alpha_f} (1-z)^{\beta_f} \; ,
\end{equation}
where $f=q,c,b,g$ and, as stated, in the case of heavy quarks 
$\mu^2=m_h^2$. Utilizing Eq.\ (\ref{eq:ansatz}) and
assuming for simplicity that $N_c=N_b=N_q$, a total of 10 free 
parameters remains to be fixed from a fit to the available 103 data 
points \cite{newdata,slddata} (after applying the $x_E$ cut mentioned above). 
The total $\chi^2$ values are 103.55 and 104.29 in NLO and LO, respectively, 
and the optimal parameters in (\ref{eq:input}) can be found in Table 1. 
%%%%%%%%
%TABLE 1 
%%%%%%%%
\begin{table}[ht]
\vspace*{-0.4cm}
\renewcommand{\arraystretch}{1.6}
\begin{center}
\begin{tabular}{|c|c|c|} \hline \hline
Parameter & LO & NLO ($\overline{\rm{MS}}$)   \\ \hline \hline
$N_q$ & $0.63$ & $0.55$ \\ \hline 
$\alpha_q$ & $0.23$ & $0.22$\\ \hline 
$\beta_q $  & $1.83$ & $2.16$ \\ \hline 
$N_g$ & $0.91$ & $2.23$ \\ \hline 
$\alpha_g$ & $1.36$ & $1.86$\\ \hline 
$\beta_g $  & $3.14$ & $3.48$ \\ \hline 
$\alpha_c$ & $-0.41$ & $-0.35$\\ \hline 
$\beta_c $  & $5.66$ & $6.06$ \\ \hline 
$\alpha_b$ & $-0.29$ & $-0.32$\\ \hline 
$\beta_b $  & $5.01$ & $5.45$ \\ \hline \hline
\end{tabular}
%\vspace*{-0.05cm} 
\caption{\sf Optimal parameters for the unpolarized fragmentation functions in
Eq.\ (\ref{eq:input}).} 
\vspace*{-0.5cm}
\end{center}
\end{table}
It should be noted that by taking into account an additional 
$4\%$ normalization uncertainty for the LEP data \cite{newdata},  
$\chi^2$ can be further reduced but without any noticeable changes in the 
distributions.

A comparison of our LO and NLO results with the data is presented in Fig.\ 1, 
where all the existing data \cite{newdata,slddata} have been 
converted\footnote{The available data sets \cite{newdata,slddata} are 
presented in terms of three different variables: $x_E$, $x_p$, and $\xi$. 
These variables are simply related to each other by $x_p=\beta x_E$ with 
$\beta=\sqrt{1-m_H^2/E_H^2}$, 
and $\xi=\ln (1/x_p)$.} to the ``format'' of Eq.(\ref{eq:cross}). 
One should note that the LO and NLO results are almost indistinguishable, 
demonstrating the perturbative stability of the process considered. 
Furthermore, there is an excellent agreement between the predictions of our 
fits and the data even in the region of ``small'' $x_E$ which has not been 
included in our analysis.

Fig.\ 2 shows our LO and NLO fragmentation functions as specified
in Eq.\ (\ref{eq:input}) and Table 1, evolved to $Q^2=100$ and $10^4$ GeV$^2$. 
As can be seen, the heavy quark fragmentation 
functions turn out to be comparable to the light quark ones for small $z$,
whereas they are suppressed for $z\gtrsim 0.3$. 
It should be also noted that our $c$ and $b$ fragmentation
functions are also in agreement with recent results from SLD \cite{slddata}
for the $c/uds$ and $b/uds$ ratios of $\Lambda$ production rates in
flavour-tagged $Z$ decays. 

In Fig.\ 3  we show our fragmentation functions for light quarks and gluons 
as functions of $Q^2$ for several fixed values of $z$. One can observe the 
importance of the QCD evolution of the fragmentation functions which we 
will use for making predictions for SIDIS at $Q^2$ values much lower than 
the ones at which the fragmentation functions were extracted. 

Finally, we investigate the contribution of our $\Lambda$ (more precisely
$\Lambda+\bar{\Lambda}$, see Eq.\ (\ref{eq:lamlambar})) fragmentation
functions $D_f^{\Lambda}(z,Q^2)$ to the momentum sum rule
\begin{equation}
\label{eq:momsum}
\sum_H \int_0^1 dz z D_f^{H}(z,Q^2) =1\;\;\;.
\end{equation}
Eq.\ (\ref{eq:momsum}) expresses the conservation of the momentum of the
fragmenting parton $f$ in the fragmentation process, i.e. each parton
$f$ will fragment with $100\%$ probability into some hadron $H$. Of course
the sum rule (\ref{eq:momsum}) should be dominated, even almost saturated,
by the fragmentation into the lightest hadrons such as $\pi$ and $K$ mesons.
Hence the contribution to (\ref{eq:momsum}) due to $D_f^{\Lambda}$ is 
expected to be rather small. Indeed we find that in LO and NLO the 
contribution of our light quark (gluon) $\Lambda+\bar{\Lambda}$ fragmentation 
functions to the momentum sum rule (\ref{eq:momsum}) only amounts 
to about $2-3\%$ ($1-2\%$).

%%%%%%%%%%%%%%%%%%%%%%%%%%%%%%%%%%%%%%%%%%%
\section{Polarized Fragmentation Functions}
%%%%%%%%%%%%%%%%%%%%%%%%%%%%%%%%%%%%%%%%%%%
%
\noindent 
Having obtained a reliable set of unpolarized fragmentation 
functions we now turn to the polarized case where unfortunately
only scarce and far less precise data are available. In fact, no data at 
all have been obtained so far using polarized beams. The
only available information comes from {\it un}polarized LEP measurements 
[10-12] profitting from the parity-violating 
electroweak $q\bar{q}Z$ coupling. 

For such measurements, done {\em{at}} the mass of the $Z$ boson 
($Z$-resonance), the cross section for the production of polarized 
hadrons can be written as \cite{burkjaf,ravi}
\begin{equation}
\label{eq:pol}
\frac{d\Delta \sigma^H}{d\Omega dx_E} = 3 \frac{\alpha^2(Q^2)}{2 s}\Big[  
g_3^H(x_E,Q^2) (1+\cos^2\theta) -  g_1^H(x_E,Q^2) \cos\theta  
+ g_L^H(x_E, Q^2) (1- \cos^2\theta) \Big] \; .
\end{equation}
If, as for the quoted experimental results [10-12], the 
cross section is integrated over the production angle $\theta$, the 
anyway small, charge suppressed contribution from $g_1^H$ drops out. 
One can then define the asymmetry
\begin{equation}
\label{eq:lepasym}
A^H = \frac{ g_3^H + g_L^H/2}{ F_1^H+F_L^H/2}
\end{equation}
which corresponds to the ``$\Lambda$-polarization'' observable measured at LEP.
The polarized structure functions $g_1^H$, $g_3^H$ and $g_L^H$ in 
(\ref{eq:pol}) and (\ref{eq:lepasym}) are given in NLO by the following 
expressions:
\begin{eqnarray}
\label{eq:g1nlo}
g_1^H(x_E,Q^2)&=& \sum_q g'_q \Bigg\{ \left[\Delta D_q^H (x_E,Q^2) +
\Delta D_{\bar q}^H (x_E,Q^2) \right] \\
&& \left.  + \frac{\alpha_s(Q^2)}{2\pi} \left[\Delta C_q^1 
\otimes (\Delta D_q^H+\Delta D_{\bar q}^H) +\Delta C_g^1 
\otimes \Delta D_g^H \right] (x_E,Q^2) \right\} \nonumber \\
\label{eq:g3nlo}
g_3^H(x_E,Q^2)&=& \sum_q \, g_q \Bigg\{ \left[\Delta D_q^H (x_E,Q^2) -
\Delta D_{\bar q}^H (x_E,Q^2) \right]  \\
&& \left.  + \frac{\alpha_s(Q^2)}{2\pi} \left[ \Delta C_q^3 
\otimes (\Delta D_q^H-\Delta D_{\bar q}^H) \right] (x_E,Q^2) 
\right\} \nonumber \\
\label{eq:glnlo}
g_L^H(x_E,Q^2)&=& \frac{\alpha_s(Q^2)}{2\pi}\sum_q \, g_q\,
\left[ \Delta C_q^L \otimes (\Delta D_q^H-\Delta D_{\bar q}^H) \right] 
(x_E,Q^2)\, ,
\end{eqnarray}
with the convolutions as already defined in (\ref{eq:convolut}).
The appropriate effective charges $g_q$ and $g'_q$ as well as the 
required spin-dependent ${\overline{\mathrm{MS}}}$ coefficients $\Delta
C_{q,g}^{1,3,L}$ in (\ref{eq:g1nlo})-(\ref{eq:glnlo}) can be found in 
Appendix B.

Note that both $g_3^H$ and $g_L^H$ in (\ref{eq:g3nlo}), (\ref{eq:glnlo}) 
are {\em{non-singlet}} structure functions, and hence 
only the {\it valence} part of the polarized fragmentation functions 
can be obtained from the available LEP data [10-12].
In addition the $\Lambda^0$'s and $\bar{\Lambda}^0$'s give contributions 
of opposite signs to the measured polarization and thus to $g_{3,L}^\Lambda$. 
Unfortunately, it is clear that the available LEP data 
[10-12], all obtained {\em on} the $Z$-resonance, 
cannot even sufficiently constrain the valence 
distributions for all the flavours, so some assumptions have to be made 
here. Obviously, even further assumptions are needed for the polarized 
gluon and sea fragmentation functions in order to have a complete set of 
fragmentation functions suitable for predictions for other processes, 
in particular for SIDIS (see Section 4).

In the present analysis the heavy flavour contributions to polarized $\Lambda$ 
production are neglected, and $u$ and $d$ fragmentation functions 
are taken to be equal. Furthermore, polarized  
``unfavoured'' distributions, i.e., $\Delta D_{\bar{u}}^{\Lambda^0} = 
\Delta D_{u}^{\bar{\Lambda}^0}$, etc., and the gluon fragmentation 
function $\Delta D_g^\Lambda$ are {\em{assumed}} to be negligible at the 
initial scale $\mu$, an assumption which of course   deserves a further
scrutiny (we will discuss the impact of choosing a different boundary condition
for the gluon fragmentation function later). The remaining spin-dependent 
quark fragmentation functions are then related to the corresponding 
unpolarized ones taken from Section 2 in the following simple way
\begin{equation}
\label{eq:polinput}
\Delta D_s^\Lambda (z,\mu^2) = z^\alpha  D_s^\Lambda (z,\mu^2)\;\;,\;\;
\Delta D_u^\Lambda (z,\mu^2) = \Delta  D_d^\Lambda (z,\mu^2) = 
N_u\, \Delta D_s^\Lambda (z,\mu^2) \;\; .
\end{equation}
They are subject to the positivity constraints (\ref{eq:positivity}),
which simply imply $\alpha>0$ and $|N_u|\le 1$. These input distributions 
are then evolved to higher $Q^2$ via the appropriate Altarelli-Parisi 
equations which are completely similar to the ones presented in 
Eqs.\ (\ref{eq:startevol})-(\ref{eq:expan}), with just all unpolarized 
quantities (like, for instance, $\hat{P}^{(T)}$) replaced by their appropriate 
polarized counterparts. For the NLO evolution one has to use for this 
purpose the spin-dependent ``time-like'' two-loop splitting functions as 
derived in \cite{tlpol} in the $\overline{\mathrm{MS}}$ scheme. 
Due to the rather limited amount of available data it does not appear
reasonable for the time being to introduce more free parameters than the two 
in Eq.~(\ref{eq:polinput}).

Within this framework we try three different scenarios 
for the polarized fragmentation functions at our low
initial scale $\mu$, to cover a rather wide range of plausible models: 

\noindent
{\bf{Scenario 1}} corresponds to the expectations from the non-relativistic 
naive quark model where only $s$-quarks can contribute to the fragmentation
processes that eventually yield a polarized $\Lambda$, even if the $\Lambda$
is formed via the decay of a heavier hyperon. We hence have $N_u=0$ in 
(\ref{eq:polinput}) for this case.

\noindent
{\bf{Scenario 2}} is based on estimates by Burkardt and 
Jaffe \cite{burkjaf,jaffe2} for the ``space-like'' DIS structure function 
$g_1^{\Lambda}$ of the $\Lambda$, predicting
sizeable negative contributions from $u$ and $d$ quarks to $g_1^{\Lambda}$ by 
analogy with the breaking of the Gourdin-Ellis-Jaffe sum rule \cite{gej}
for the proton's $g_1^p$. Assuming that such features also carry over
to the ``time-like'' case \cite{jaffe2}, we simply impose $N_u=-0.20$ 
(see also \cite{bravar}).

\noindent
{\bf{Scenario 3:}} All the polarized fragmentation functions are assumed 
to be equal here, i.e. $N_u=1$, contrary to the expectation  
of the non-relativistic quark model used in scen.\ 1. This rather ``extreme'' 
scenario might be realistic if, for instance, there are sizeable 
contributions to polarized $\Lambda$ production from decays of 
heavier hyperons who have inherited the polarization of originally 
produced $u$ and $d$ quarks.

%
%%%%%%%%
%TABLE 2 
%%%%%%%%
\begin{table}[ht]
\vspace*{-0.4cm}
\renewcommand{\arraystretch}{1.6}
\begin{center}
\begin{tabular}{|c|c|c|c|c|c|c|} \hline \hline
Parameter & \multicolumn{3}{c|}{LO} & \multicolumn{3}{c|}
{NLO ($\overline{\mathrm{MS}}$)}  \\ \hline \hline
& scen. 1 & scen. 2 & scen. 3 & scen. 1 & scen. 2 & scen. 3 \\ \hline  
$N_u$& 0 &$-0.2 $&$ 1.0 $&$ 0 $&$ -0.2$&$ 1.0 $ \\ \hline 
$\alpha$ &$ 0.62  $&$ 0.27 $&$ 1.66 $&$ 0.44 $&$  0.13 $&$ 1.33 $\\ \hline 
\hline
\end{tabular}
%\vspace*{-0.05cm} 
\caption{\sf Resulting optimal LO and NLO fit parameters as introduced in 
(\ref{eq:polinput}) for the three different scenarios described in the text.} 
\vspace*{-0.5cm}
\end{center}
\end{table}
Our results for the asymmetry $A^{\Lambda}$ in (\ref{eq:lepasym}) within 
the three different scenarios are compared to the available LEP 
data [10-12] in Fig.\ 4. The optimal parameters in 
(\ref{eq:polinput}) for the three models can be found in Table 2. As can 
be seen, the best agreement with the data is obtained within the 
(naively) most unlikely scen.\ 3. The differences occur mainly in the region 
of large $x_E$, where scen.\ 1 and 2 cannot fully account for the rather large 
observed polarization. It turns out that this is a consequence of the 
assumed $SU(3)_f$ symmetry for the {\it{un}}polarized fragmentation functions,
and of the positivity constraints (\ref{eq:positivity}): for instance, 
$SU(3)_f$ symmetry of the $D_q^{\Lambda}$ implies in the case of scen.\ 1 that 
the asymmetry at large $x_E$  behaves asymptotically roughly like 
$-\Delta D_s^\Lambda / 3D_s^\Lambda$. Thus, even when saturating the 
positivity constraint (\ref{eq:positivity}) at around $x_E=0.5$ it is not 
possible to obtain a polarization as large as the one required by the 
ALEPH and OPAL data \cite{aleph,opal}. 
We note that the assumed $SU(3)_f$ symmetry for the unpolarized 
fragmentation functions could of course be broken. It is clear at this
point that further information on the polarized {\em and} the 
unpolarized $\Lambda$ fragmentation functions is needed, which could be
provided by future precise SIDIS measurements.

Finally, in Fig.\ 5 we show the LO and NLO partonic fragmentation 
asymmetries for each flavour distribution separately, 
i.e. $A_f\equiv\Delta D^{\Lambda}_f/D^{\Lambda}_f$. A positive polarized 
gluon fragmentation function has built up in the $Q^2$-evolution in spite 
of the vanishing input at $\mu^2$. In order to analyze the effect of 
imposing a different boundary condition for the polarized gluon fragmentation 
function, we include in Fig.\ 5 also the results of a LO fit similar to the 
one performed within scenario 1, but now using the maximally allowed 
polarized gluon input $\Delta D_g(\mu^2)=D_g(\mu^2)$
instead of $\Delta D_g(\mu^2)=0$.
Besides the expected 
result of having now a larger gluon polarization at $Q^2=10$ GeV$^2$, an 
important enhancement of the $u$ and $d$ distributions (which are practically 
vanishing in the original scenario 1) can be observed, which is due to the 
perturbative generation of sea by polarized gluons in the course of the 
evolution. In fact, at small values of $z$, the $u$ and $d$ distributions 
become even larger than the ones of scenario 3. Obviously, only different 
combined further measurements, like in $e^+ e^- $ annihilation and SIDIS with 
polarized beams, will be capable of determining the gluon (and also the sea) 
fragmentation function more precisely.

%%%%%%%%%%%%%%%%%%%%%%%%%%%%%%%%%%%%%%%
\section{$\Lambda$ Production in SIDIS}
%%%%%%%%%%%%%%%%%%%%%%%%%%%%%%%%%%%%%%%
\noindent
Equipped with various sets of polarized fragmentation 
functions, let us now turn to the SIDIS process $e N \rightarrow e' H X$ 
which should be very well suited to give further information on fragmentation 
functions. In this case, the cross section is proportional to a combination of 
both the parton distributions of the nucleon $N$ and the fragmentation 
functions for the hadron $H$. The latter thus automatically appear in a 
constellation different from the one probed in $e^+ e^-$ annihilation. 
 
In the particular case where both nucleon and hadron are unpolarized, the 
cross section can be written in a way similar to the fully inclusive 
DIS case \cite{altarelli,fupe,graudenz}:
\begin{equation}
\label{eq:sidis}
\frac{d\sigma^H}{dx\, dy\, dz_H} = 
\frac{2\, \pi\alpha^2}{Q^2} 
\left[ \frac{(1+(1-y)^2)}{y} 2\, F_1^{N/H}(x,z_H,Q^2) + 
\frac{2 (1-y)}{y} F_L^{N/H}(x,z_H,Q^2) \right] \; \; ,
\end{equation}
with $x$ and $y$ denoting the usual DIS scaling variables $(Q^2=s x y)$, and
where \cite{altarelli,fupe} $z_H\equiv p_H\cdot p_N/p_N\cdot q$
with an obvious notation of the four-momenta, and with $-q^2\equiv Q^2$.
Strictly speaking, Eq.\ (\ref{eq:sidis}) and the variable $z_H$ only apply 
to hadron production in the current fragmentation region. In this work, 
we will effectively eliminate the target fragmentation region by implementing 
a cut $x_F>0$ on the Feynman-variable representing the fractional 
longitudinal c.m.s.\ momentum. Target fragmentation could be accounted 
for by transforming to the variable [29-32] 
\begin{equation} 
\label{eq:zdef}
z_H \rightarrow z \equiv \frac{E_H}{E_N (1-x)} \; \; ,
\end{equation}
the energies $E_H$, $E_N$ defined in the c.m.s. frame of the nucleon and the
virtual photon, and by introducing the so-called ``fracture functions'' 
\cite{veneziano}. The inclusion of the latter is beyond the scope of this 
analysis \cite{ellis} and anyway not relevant numerically due to the cut
on $x_F$. The variable $z$ in (\ref{eq:zdef}) 
is also better suited for dealing with corrections due to the finite target 
mass $M_H$. As will be demonstrated below, it is not always 
justified to neglect these. Our predictions for $\Lambda$ production in 
SIDIS will therefore be made using the variable $z$. The NLO 
corrections to $F_1^{N/H}$ and $F_L^{N/H}$ in
(\ref{eq:sidis}) can, however, be expressed much more conveniently 
in terms of $z_H$ (see Appendix C). The transformation from $z_H$ to
$z$ is straightforward [29-32].

The structure functions $F_1^{N/H}$ and $F_L^{N/H}$ in (\ref{eq:sidis})
are given at NLO by 
\begin{eqnarray}
\label{eq:f1sidis}
2\, F_{1}^{N/H}(x,z_H,Q^2) &=& \sum_{q,\overline{q}} e_q^2 \left
\{  q (x,Q^2)  D^H_q (z_H,Q^2) +\frac{\alpha_s(Q^2)}{2\pi}\bigg[ q
\otimes  C^1_{qq} \otimes D^H_q \right.  \nonumber \\
&+& q  \otimes  C^1_{gq} \otimes D^H_g+  g  \otimes
C^1_{qg} \otimes D^H_q \bigg] (x,z_H,Q^2) \Bigg\} \\
\label{eq:flsidis}
F_{L}^{N/H}(x,z_H,Q^2) &=& \frac{\alpha_s(Q^2)}{2\pi} 
\sum_{q,\overline{q}} e_q^2 \bigg[ q 
\otimes  C^L_{qq} \otimes D^H_q   \nonumber \\
&+& q  \otimes  C^L_{gq} \otimes D^H_g+  g  \otimes
C^L_{qg} \otimes D^H_q \bigg] (x,z_H,Q^2) \;\;, 
\end {eqnarray}
with the NLO coefficient functions $C^{1,L}_{ij}$ 
\cite{altarelli,fupe,graudenz} collected in Appendix C. 
 
As already mentioned in the Introduction, three other possible cross
sections can be defined when the polarization of the lepton, the initial 
nucleon and the hadron are taken into account. 
If both nucleon and hadron are polarized and the lepton is unpolarized, 
the expression is similar to Eqs.\ (\ref{eq:sidis})-(\ref{eq:flsidis}) 
above with, however, the unpolarized parton distributions {\em and} the 
fragmentation functions to be replaced by their polarized counterparts. 
Obviously, one also has to adapt the coefficient functions 
to this case: $C^{1,L}_{ij}\rightarrow \Delta C^{1,L,NH}_{ij}$.
The relevant expressions can again be found in Appendix C. 
In the case that the lepton and either the nucleon {\em{or}} 
the hadron are polarized, the expression for the cross 
section is given as in the fully inclusive case by a single structure 
function $g_{1}^{N/H}(x,z_H,Q^2)$: 
\begin{equation}
\label{eq:sidispol}
\frac{d\Delta\sigma^H}{dx\, dy\, dz_H} = \frac{4 \pi\alpha^2}{Q^2} 
(2-y)\, g_1^{N/H}(x,z_H,Q^2)  \; .
\end{equation}
To NLO, $g_1^{N/H}$ can be written as \cite{singlepol,doublepol,thesis}
\begin{eqnarray}
\label{eq:pol1} 
2\, g_{1}^{N/H}(x,z_H,Q^2) &=& \sum_{q,\overline{q}} e_q^2 \Bigg\{
(\Delta)q (x,Q^2) (\Delta)D^H_q (z_H,Q^2) \nonumber \\
&+& \frac{\alpha_s(Q^2)}{2\pi} \bigg[ (\Delta)q \otimes \Delta C^i_{qq} 
\otimes (\Delta)D^H_q + (\Delta)q  \otimes  \Delta C^i_{gq} 
\otimes (\Delta)D^H_g \nonumber  \\
&+& (\Delta) g  \otimes \Delta C^i_{qg} \otimes (\Delta)D^H_q \bigg] 
(x,z_H,Q^2)\Bigg\} \;\;,
\end{eqnarray}
the position of the $\Delta$ and the index $i=N,\,H$ depending on which 
particle ($N$ or $H$) is polarized. Again, all NLO $\overline{\mathrm{MS}}$ 
coefficient functions $\Delta C_{jk}^i$ are collected in Appendix C.

Let us first turn to the entirely unpolarized case as defined in 
Eq.\ (\ref{eq:sidis}), which could prove invaluable for obtaining a 
flavour separation of fragmentation functions not provided by the 
SIA data. Unfortunately, only three measurements of this cross section 
exist up to now \cite{h1dis}, with still rather large experimental 
uncertainties. It is nevertheless worth comparing our predictions 
to the available data in order to test our proposed fragmentation functions
in a process other than SIA.

The original experimental results \cite{h1dis} are compiled in Fig.\ 6a, 
where the data are plotted in terms of the Feynman variable $x_F$. As 
can be observed, differences between results at different values of 
the $\gamma^* p\;$ c.m.s.\ energy $W$ are much larger than 
expected from the scale dependence, especially at small $x_F$. 
The reason for this is simple and corresponds to the 
fact that the variable $x_F$ is not the scaling variable of this 
process, i.e. the argument of the fragmentation functions in 
Eq.\ (\ref{eq:sidis}). As already mentioned, Eq.\ (\ref{eq:sidis}) 
should be expressed in terms of $z$ in (\ref{eq:zdef}) \cite{graudenz}.
At LO, $z$ coincides with $x_F$ and also with the variable $z_H$ 
introduced in (\ref{eq:sidis}) \cite{altarelli,fupe}, {\em if} the mass of 
the $\Lambda$ is neglected. However, it again turns out that, 
as in the case of SIA, finite-mass effects introduced by the function 
$\beta= \sqrt{1- 4 M_\Lambda^2/(z W)^2}$ become relevant at small $z$ for 
the low-$W$ experiments. The other two variables are at LO given in terms 
of $\beta$ and $z$ by $x_F=\beta\, z$ and $z_H=(1+\beta)/2\, z$. 
In Fig.\ 6b we show the same data as in Fig.\ 6a, but converted to the 
variable $z$, where now a much better agreement between different 
experimental data and also with our LO predictions plotted for 
three different typical scales can be observed. It should be noticed that the 
H1 data \cite{h1dis} were obtained with an integrated luminosity of only 
1.3 pb$^{-1}$, so a more dedicated measurement in the future will be very 
helpful in determining the unpolarized $\Lambda$ fragmentation functions 
more precisely.

The most interesting observable with respect to the determination
of the {\em polarized} $\Lambda$ fragmentation functions is of course the 
asymmetry for the production of polarized $\Lambda$'s from an 
{\em un}polarized proton, defined by $A^{\Lambda}\equiv g_1^{p/\Lambda}/
F_1^{p/\Lambda}$ \cite{jaffe2} with $g_1^{p/\Lambda}$ given by
(\ref{eq:pol1}) with $i=H$. In Fig.\ 7a, we show our LO and NLO 
predictions for HERA with polarized electrons and {\it{un}}polarized 
protons using the GRV parton distributions \cite{grv}, integrated over 
the measurable range $0.1 \leq z \leq 1$. The values for $Q^2$ that 
correspond to each $x$-bin have 
been chosen as in \cite{desh}. Good perturbative stability of 
the process is found. As can be seen, the results obtained using the 
three distinct scenarios for polarized fragmentation functions turn 
out to be completely different. Since the asymmetry at small 
$x$ is determined by the proton's sea quarks, its behaviour can 
be easily understood: 
in scen.\ 1 only $s$ quarks fragment into polarized $\Lambda$'s, 
giving an asymmetry which is positive but about three times 
smaller than the one of scen.\ 3 where all the flavours contribute. 
In the case of scen.\ 2 the positive contribution from the $s$-quark 
fragmentation is cancelled by a negative one from $u$ and $d$, resulting 
in an almost vanishing asymmetry. The interpretation is similar for 
the region of large $x$, where only the contribution involving $u_{v}$ 
is sizeable and the asymmetry 
asymptotically goes to $\int \,dz\Delta D_u^\Lambda / \int \,dz 
D_u^\Lambda$ for each scenario.

We have included in Fig.\ 7a also the expected statistical errors 
for HERA, computed assuming an integrated luminosity of 
$500$ pb$^{-1}$ and a realistic value of $\epsilon=0.1$ for the 
efficiency of $\Lambda$ detection \cite{albert}. Comparing the asymmetries 
and the error bars in Fig.\ 7a one concludes that a measurement of 
$A^{\Lambda}$ at small $x$ would allow a discrimination 
between different conceivable scenarios for polarized fragmentation functions.
Fig.\ 7b shows our results vs.\ $z$ for fixed $x=5.6 \cdot 10^{-4}$. Again, 
very different asymmetries are found for the three scenarios. 
In this plot we also include the expectation from scenario 1 with a 
maximal gluon polarization (see Sec.\ 3) which, as expected, predicts a 
larger asymmetry at small $z$ (comparable to the one of scenario 3) 
mainly due to the contribution of the radiated sea. 

In Fig.\ 8 we show the same observable for the case of HERMES, where the 
$Q^2$ values were chosen as for the inclusive DIS measurements by HERMES
\cite{hermes}. This fixed target experiment analyses a different kinematical 
region of both larger $x$ and $z$ ($z>0.3$) and hence could provide 
complementary information. As can be seen by comparing Figs.\ 7b and 8b, 
the asymmetry for scenario 3 shows a similar behaviour for both experiments, 
which is expected as all the fragmentation functions are equal and the 
parton distributions cancel in the ratio (at LO). However, the asymmetries 
predicted by scenarios 1 and 2 change quite a bit when going from HERA
collider to fixed target energies due to the fact that for the values of $x$
probed at HERMES the contributions from the valence distributions dominate.
Thus more weight is given to the $D_{u,d}^{\Lambda}$ 
fragmentation functions, and the
contributions involving $D_s^{\Lambda}$ are suppressed, 
in contrast to the situation 
for the small $x$-region to be explored by HERA. Again we show in Fig.\ 8b 
also the results for scenario 1 with a maximally saturated gluon 
fragmentation function at the input scale which leads to
results hardly different from the standard one since the sea contribution is 
negligible at large $z$.

Finally, the particular case of both target and hadron being 
polarized was originally proposed as a very good way to obtain 
the $\Delta s$ distribution \cite{luma}. 
The underlying assumption here was that only the fragmentation function 
$\Delta D_s^{\Lambda}$ is sizeable (as realized, e.g. in our scenario 1), and 
that therefore the only contribution to the polarized cross section has to be 
proportional to $\Delta s \Delta D_s^{\Lambda}$. In order to analyze the 
sensitivity of the corresponding asymmetry to $\Delta s$, we compute it 
using the two different GRSV sets of polarized parton densities of the 
proton \cite{grsv}, which mainly differ in the strange distribution:
the so-called ``standard'' set assumes an unbroken $SU(3)_f$ symmetric
sea, whereas in the ``valence'' scenario the sea is maximally broken
and the resulting strange quark density is quite small.

The results for HERA are shown in Fig.\ 9. 
Unfortunately -- and not unexpectedly -- it turns out that the 
differences in the asymmetry resulting from our 
different models for polarized $\Lambda$ fragmentation are far larger 
than the ones due to employing different polarized proton strange densities.
In addition, a distinction between different $\Delta s$ would remain 
elusive even if the spin-dependent $\Lambda$ fragmentation functions
were known to good accuracy, as can be seen from the error bars
in Fig.\ 9 which were obtained using the same parameters as before.
%
%%%%%%%%%%%%%%%%%%%%%%%%%%%%%%%%%
\section{Summary and Conclusions}
%%%%%%%%%%%%%%%%%%%%%%%%%%%%%%%%%
\noindent
We have performed a detailed QCD analysis of the production of $\Lambda$ 
baryons in $e^+e^-$ annihilation and semi-inclusive deep-inelastic 
scattering.

Working within the framework of the radiative parton model, our starting 
point has been a fit to unpolarized data for $\Lambda$ production taken
in $e^+ e^-$ annihilation, yielding a set of realistic unpolarized 
fragmentation functions for the $\Lambda$. We have then made simple 
assumptions for the relation between the spin-dependent and the unpolarized 
$\Lambda$ fragmentation functions at the input scale for the $Q^2$-evolution. 
Taking into account the sparse LEP data on the polarization of $\Lambda's$
produced on the $Z$-resonance,
we were able to set up three distinct ``toy scenarios'' for the spin-dependent
$\Lambda$ fragmentation functions, to be used for predictions for future
experiments. We emphasize that our proposed sets can by
no means cover all the allowed possibilities for the polarized fragmentation 
functions, the main reason being that the LEP data are only sensitive
to the valence part of the polarized fragmentation functions. Thus, there are 
still big uncertainties related to the ``unfavoured'' quark and gluon 
fragmentation functions, making further measurements in other processes
indispensable.

Under these premises, we have studied $\Lambda$ production in 
semi-inclusive deep-inelastic scattering. Existing data for the production 
of unpolarized $\Lambda$'s are well described by our fragmentation functions
determined from the $e^+ e^-$ annihilation data. Turning to spin transfer 
asymmetries sensitive to the longitudinal polarization 
of the produced $\Lambda$'s,
we have considered both $\vec{e}p\rightarrow \vec{\Lambda}X$ and 
$e\vec{p}\rightarrow \vec{\Lambda}X$ scattering. It turns out that in the first
case SIDIS measurements at HERA (with spin-rotators in front of the H1 and 
ZEUS detectors) and at HERMES should be particularly well suited to 
yield further information on the $\Delta D_f^{\Lambda}$: differences 
between the asymmetries obtained when using different sets of 
$\Delta D_f^{\Lambda}$ are usually larger than the expected statistical 
errors. In contrast to this, having a polarized proton target (or beam) does 
not appear beneficial as far as $\Lambda$ production is concerned.   

A {\sc{Fortran}} package containing our unpolarized and polarized LO and NLO 
$\Lambda$ fragmentation functions can be obtained by electronic mail from  
Daniel.Deflorian@cern.ch, Marco.Stratmann@durham.ac.uk, or 
Werner.Vogelsang@cern.ch upon request.

\section*{Acknowledgements}
The work of one of us (DdF) was partially supported by the World Laboratory.

%%%%%%%%%%%%%%%%%%%%%%%%%%%%%%%%%%%%%%
\section*{Appendix A: Unpolarized SIA}
%%%%%%%%%%%%%%%%%%%%%%%%%%%%%%%%%%%%%%
\renewcommand{\theequation}{{\rm{A}}.\arabic{equation}}
\setcounter{equation}{0}
\noindent
The NLO ($\overline{\mathrm{MS}}$) coefficients $C_{q,g}^{1,L}$ in
(\ref{eq:f1nlo}) and (\ref{eq:flnlo}) are given by \cite{altarelli,fupe}:
\begin{eqnarray}
\label{eq:appa1}
C_q^1(z)&=& C_F \left[ (1+z^2) 
\left(\frac{\ln (1-z)}{1-z}\right)_+ - \frac{3}{2} \frac{1}{(1-z)_+} 
\right. \\ 
&& \left.  + 2\, \frac{1+z^2}{1-z} \ln z+ 
\frac{3}{2} (1-z) + \left(\frac{2}{3} \pi^2 - 
\frac{9}{2}\right) \delta(1-z) \right] \nonumber \\
C_g^1(z)&=&2\, C_F \left[ \frac{1+(1-z)^2}{z} \, \ln \left( z^2(1-z) \right) - 
2\, \frac{1-z}{z} \right]  \\
\label{eq:appa7}
C_q^L(z)&=& C_F  \\
\label{eq:appa2}
C_g^L(z)&=&4 \, C_F \, \frac{(1-z)}{z}  
\end{eqnarray} 
with $C_F=4/3$. Note that in the expressions for $C_{q,g}^1$ we 
have taken the factorization scales for the final-state mass singularities
to be equal to the hard scale $Q$ of the process, as we did in all our
numerical applications. The ``+''-prescription is defined as usual by
\begin{equation}
\label{eq:plusdist}
\int_0^1 dz f(z) \left( g(z) \right)_+ \equiv 
\int_0^1 dz \left[ f(z)-f(1)\right] g(z) \:\:.
\end{equation}

The electroweak charges in (\ref{eq:cross})-(\ref{eq:flnlo}) are given by
\begin{equation}
\hat{e}_q^2= e_q^2 - 2 e_q \chi_1 (Q^2) V_e V_q + \chi_2 (Q^2) (1+V_e^2) 
(1+ V_q^2)
\end{equation}
where
\begin{eqnarray}
\label{eq:appa3}
\chi_1 (s) &=&\frac{1}{16\, \sin^2\Theta_W \cos^2\Theta_W} 
\frac{s \left(s-M_Z^2\right)}{(s-M_Z^2)^2+\Gamma_Z^2 M_Z^2} \nonumber \\
\label{eq:appa4}
\chi_2 (s) &=&\frac{1}{256\, \sin^4\Theta_W \cos^4\Theta_W} 
\frac{s^2}{(s-M_Z^2)^2+\Gamma_Z^2 M_Z^2} \; .
\end{eqnarray}
Here $e_q$ is the fractional electromagnetic quark charge, and $M_Z$ 
and $\Gamma_Z$ are the mass and the decay width of the $Z$ boson, 
respectively. The other electroweak couplings are given in terms of the 
Weinberg angle $\Theta_W$ by
\begin{eqnarray}
\label{eq:appa5}
V_e &=&-1 + 4 \sin^2\Theta_W \nonumber \\
V_u &=&+1-\frac{8}{3} \sin^2\Theta_W \nonumber \\  
V_d &=&-1+\frac{4}{3} \sin^2\Theta_W
\label{eq:appa6} \; \; .
\end{eqnarray}
%
%%%%%%%%%%%%%%%%%%%%%%%%%%%%%%%%%%%%
\section*{Appendix B: Polarized SIA}
%%%%%%%%%%%%%%%%%%%%%%%%%%%%%%%%%%%%
\renewcommand{\theequation}{{\rm{B}}.\arabic{equation}}
\setcounter{equation}{0}
The NLO $\overline{\mathrm{MS}}$ coefficients $\Delta C_{q,g}^{1,3,L}$ in 
(\ref{eq:g1nlo})-(\ref{eq:glnlo}) read:
\begin{eqnarray}
\label{eq:appb1}
\Delta C_q^1(z)&=& C_q^1(z) - C_F \left[ 1-z \right] \\
\Delta C_g^1(z)&=&2 C_F \Big[ (2-z) \, 
\ln \left( z^2(1-z)\right)- 4 + 3\, z \Big] \\
\label{eq:appb2}
\Delta C_q^3(z)&=&  C_q^1(z)   \\
\label{eq:appb3}
\Delta C_q^L(z) &=&  C_q^L(z) \;\;\;,
\end{eqnarray} 
where the effective charges {\em{on}} the $Z$-resonance are
given by \cite{burkjaf}
\begin{eqnarray}
g_q  &=& \chi_2 (M_Z^2) A_q\, V_q (1+V_e^2) \\
g'_q &=& 2\,\chi_2 (M_Z^2) V_e (1+V_q^2) \;\; ,
\end{eqnarray}
where $A_u=-A_d=1$ and $V_e,$ $V_q$, $\chi_2 (s)$ have already 
been defined in (\ref{eq:appa3}), (\ref{eq:appa6}). The structure 
functions $g_3^H$ and $g_L^H$ in Eqs.\ (\ref{eq:g3nlo}), (\ref{eq:glnlo}) 
are purely non-singlet and therefore do not receive a gluonic correction.

One should note that the NLO quark corrections for the unpolarized case, 
see Eqs.\ (\ref{eq:appa1}) and (\ref{eq:appa7}), and the ones for the 
spin-dependent parity violating structure functions $g^H_3$ and $g^H_L$ in 
(\ref{eq:appb2}), (\ref{eq:appb3}) are identical, which results from 
identical tensorial structures at the parton level. The expressions
for $\Delta C_q^1(z)$ and $\Delta C_g^1(z)$ in the $\overline{\mathrm{MS}}$
scheme were already derived in \cite{doublepol,tlpol}. The difference 
$\Delta C_q^1(z)-C_q^1(z)$ in (\ref{eq:appb1}) is independent of the 
regularization prescription chosen and coincides with the one 
found in \cite{ravi} by using off-shell gluons to regularize the collinear
singularities. 
 
%%%%%%%%%%%%%%%%%%%%%%%%%%%%%%%%%%%%%%%%%%%%%%%%%%%%%%%%%%%%%%%%%%%%%%%%%%%%
\section*{Appendix C: Unpolarized and Polarized SIDIS Coefficient Functions}
%%%%%%%%%%%%%%%%%%%%%%%%%%%%%%%%%%%%%%%%%%%%%%%%%%%%%%%%%%%%%%%%%%%%%%%%%%%%
%
\setcounter{equation}{0}
\renewcommand{\theequation}{\rm{C}.\arabic{equation}}
Here we list all unpolarized and polarized NLO ($\overline{\mathrm{MS}}$)
coefficients $(\Delta)C_{\cdots}^{\cdots}$ for SIDIS as introduced in 
Section 4. To keep the expressions as short as possible it is convenient to
define the following abbreviations
\begin{eqnarray}
\nonumber
\label{sidiseq6}
\tilde{P}_{qq} (\xi)= \frac{1+\xi^2}{(1-\xi)_+} &+& \frac{3}{2} 
\delta (1-\xi)\;\;,\nonumber\\
\tilde{P}_{gq}(\xi)=\frac{1+(1-\xi)^2}{\xi}\;\; &,& \;\; 
\Delta \tilde{P}_{gq}(\xi)=\frac{1-(1-\xi)^2}{\xi}=2-\xi\;\;,\nonumber\\
\nonumber
\tilde{P}_{qg}(\xi)= \xi^2+(1-\xi)^2\;\; &,& \;\; 
\Delta \tilde{P}_{qg}(\xi)=\xi^2-(1-\xi)^2=2\xi -1 \;\; , \\
L_1(\xi)=(1+\xi^2)\left(\frac{\ln (1-\xi)}{1-\xi}\right)_+\;\; &,& \;\;
L_2(\xi)=\frac{1+\xi^2}{1-\xi}\ln \xi\;\;.
\end{eqnarray}
Note that in what follows we always suppress the argument $(x,z)$ of the 
coefficient functions. $M$ and $M_F$ denote
the factorization scales for initial and final state mass singularities,
respectively. Note that for all our numerical calculations we have chosen 
as usual $M=M_F=Q$.
All results presented here are given in the $\overline{\mathrm{MS}}$
scheme, and in case of the spin-dependent coefficients $\Delta C_{qq}^i$
the additional finite subtractions that are required when using the 
$\gamma_5$ prescription of \cite{hvbm}, have been performed along the lines 
discussed in \cite{slsplit,tlpol}.

\noindent
{\bf{Coefficients for $eN\rightarrow e'H X$: \cite{fupe}}}
\begin{eqnarray}
\label{sidiseq8}
\nonumber
C_{qq}^1 &=& C_F\Bigg[ -8\delta(1-x)\delta(1-z)+\\
\nonumber
&&
\delta(1-x) \left[ \tilde{P}_{qq}(z) \ln\frac{Q^2}{M_F^2} +
 L_1(z)+L_2(z)+(1-z)\right] + \\
\nonumber
&&
\delta(1-z) \left[ \tilde{P}_{qq}(x) \ln\frac{Q^2}{M^2} +
L_1(x)-L_2(x)+(1-x)\right]+\\
&& 2\frac{1}{(1-x)_+}\frac{1}{(1-z)_+}- \frac{1+z}{(1-x)_+}-
\frac{1+x}{(1-z)_+}+2(1+xz) \Bigg]\\
\label{sidiseq9}
\nonumber
C_{gq}^1 &=& C_F\Bigg[ \tilde{P}_{gq}(z) \left(\delta (1-x) \ln\left(
\frac{Q^2}{M_F^2}z(1-z)\right)+\frac{1}{(1-x)_+}\right)+ \\
&&  z \delta(1-x)+2(1+x-xz)-\frac{1+x}{z}\Bigg]\\
\label{sidiseq10}
\nonumber
C_{qg}^1 &=& \frac{1}{2} \Bigg[ \delta (1-z) \left[\tilde{P}_{qg}(x)
\ln\left(\frac{Q^2}{M^2} \frac{1-x}{x}\right) +2x(1-x)\right]+ \\
&& \tilde{P}_{qg}(x) \left\{ \frac{1}{(1-z)_+}+\frac{1}{z}-2\right\} \Bigg]\\
C_{qq}^L &=& 4 C_F x z \\
C_{gq}^L &=& 4 C_F x (1-z) \\
C_{qg}^L &=& 4 x(1-x) 
\end{eqnarray}

\noindent
{\bf{Coefficients for $\vec{e}\vec{N}\rightarrow e'HX$}:}
\begin{eqnarray}
\label{sidiseq12}
\Delta C_{qq}^N &=& C_{qq}^1 - 2 C_F (1-x) (1-z) \\ 
\label{sidiseq13}
\Delta C_{gq}^N &=& C_{gq}^1 - 2 C_F z (1-x)\\
\label{sidiseq14}
\nonumber
\Delta C_{qg}^N &=& \frac{1}{2} \Bigg( \delta(1-z) \left[
\Delta \tilde{P}_{qg}(x)\ln\left(\frac{Q^2}{M^2} \frac{1-x}{x}\right) 
+2(1-x)\right]+\\
&& \Delta \tilde{P}_{qg}(x)\left[\frac{1}{(1-z)_+}+\frac{1}{z}-2\right]\Bigg)
\end{eqnarray}

\noindent
{\bf{Coefficients for $\vec{e}N \rightarrow e'\vec{H}X$}:}
\begin{eqnarray}
\label{sidiseq16}
\Delta C_{qq}^H &=&  \Delta C_{qq}^N + 2 C_F (1-z) \delta (1-x)\\
\label{sidiseq17}
\nonumber
\Delta C_{gq}^H &=& C_F \Bigg\{\Delta \tilde{P}_{gq}(z)\left[\delta(1-x)
\ln\left(\frac{Q^2}{M_F^2}z(1-z)\right)+\frac{1}{(1-x)_+}\right]- \\
&&  2(1-z) \delta(1-x)-2(1+x-z)+\frac{1+x}{z}\Bigg\}\\
\label{sidiseq18}
\Delta C_{qg}^H &=& C_{qg}^1 - \tilde{P}_{qg}(x) \frac{1-z}{z} 
\end{eqnarray}

\noindent
{\bf{Coefficients for $e\vec{N}\rightarrow e'\vec{H}X$}:}
\begin{eqnarray}
\label{sidiseq20}
\Delta C_{qq}^{1,NH} &=& C_{qq}^1 + 2C_F (1-z) \delta(1-x)\\
\label{sidiseq21}
\Delta C_{gq}^{1,NH} &=& \Delta C_{gq}^H - 2 C_F z (1-x) \\
\label{sidiseq22}
\Delta C_{qg}^{1,NH} &=& \Delta C_{qg}^N - \Delta \tilde{P}_{qg}(x) 
\frac{1-z}{z} \;\; \\
\Delta C_{qq}^{L,NH} &=& C_{qq}^L \\
\Delta C_{gq}^{L,NH} &=& -4 C_F x (1-z) \\
\Delta C_{qg}^{L,NH} &=& 0 \;\;\;.
\end{eqnarray}

We note that all our results for the spin-dependent coefficients in 
(C.8)-(C.19) coincide with the ones presented in \cite{thesis} and also
fully agree with the results of \cite{singlepol,doublepol} if one carefully 
disentangles in these papers the contributions from the current and the 
target fragmentation regions (in the same way our unpolarized results 
in (C.2)-(C.7) are in agreement with ref.\ \cite{graudenz}). In addition, 
one has to account for the slightly different factorization scheme used 
in ref.\ \cite{singlepol}.

Finally let us show how to deal with the ``+'' - distributions appearing in
the expressions above. The ``+'' - distribution was already defined 
in Eq.\ (\ref{eq:plusdist}) in Appendix A. In practice, however, the lower 
limit of the integration in (\ref{eq:plusdist}) is different from zero, 
hence the distributions have to be modified according to \cite{hinchliffe}:
\begin{eqnarray}
\nonumber
\label{sidiseq24}
\frac{1}{(1-\xi)_+} &=& \frac{1}{(1-\xi)_A} +\ln (1-A)\delta (1-\xi)\;,\\
\left(\frac{\ln 1-\xi}{1-\xi}\right)_+ &=& 
\left(\frac{\ln 1-\xi}{1-\xi}\right)_A
+\frac{1}{2} \ln^2(1-A) \delta (1-\xi)  \;\; ,
\end{eqnarray}
where $(\,)_A$ is defined as in (\ref{eq:plusdist}) but with the lower 
integration limit replaced by $A$.
In addition, in the coefficients $(\Delta) C_{qq}^i$ listed above also
double ``+'' - distributions appear, which can be defined 
in analogy with Eq.\ (\ref{eq:plusdist}) by
\begin{equation}
\label{sidiseq25}
\int_0^1 dx\int_0^1 dz \frac{f(x,z)}{(1-x)_+ (1-z)_+} \equiv
\int_0^1\int_0^1 dx dz\; \frac{f(x,z)-f(1,z)-f(x,1)+f(1,1)}{(1-x)(1-z)}\;\;.
\end{equation}
Again, in practice the lower integration limits are both different
from zero, say, $A$ for the $x$ integration and $B$ for the $z$ integration,
and the distribution defined above can be rewritten as
\begin{eqnarray}
\label{sidiseq26}
\nonumber
\frac{1}{(1-x)_+(1-z)_+}= \frac{1}{(1-x)_A(1-z)_B} +
\frac{1}{(1-x)_A} \ln (1-B)\delta (1-z) +\\
\frac{1}{(1-z)_B} \ln (1-A)\delta (1-x) +
\ln (1-A) \ln (1-B) \delta(1-x)\delta(1-z)\;\;.
\end{eqnarray}
%
%%%%%%%%%%%%%%%%%%%%%%%%%%%%%

%
%\newpage
%
%%%%%%%%%%%%%%%%%%%%
% FIGURE CAPTIONS
%%%%%%%%%%%%%%%%%%%%
\section*{Figure Captions}
\begin{description}
\item[Fig.\ 1] Comparison of our LO and NLO results for
$(1/\sigma_{tot}) d\sigma/dx_E$ according to 
Eq.\ (\ref{eq:cross}) with all available data on 
unpolarized $\Lambda$ production in $e^+ e^-$ annihilation 
\cite{newdata,slddata}. Note that only data points with 
$x_E\ge 0.1$ have been included in our fit (see text).
\item[Fig.\ 2] $z$-dependence of our LO and NLO fragmentation 
functions as specified in Eq.\ (\ref{eq:input}) and Table 1 at $Q^2=100$ and 
$Q^2=10^4$ GeV$^2$.
\item[Fig.\ 3] $Q^2$-dependence of our LO and NLO $q$ and $g$ fragmentation 
functions at fixed $z=0.05,\, 0.1,\, 0.5$.
\item[Fig.\ 4] Comparison of LEP data [10-12] and our LO 
and NLO results for the asymmetry $A^{\Lambda}$ in (\ref{eq:lepasym}), 
using the three different scenarios as described in the text.
\item[Fig.\ 5] LO and NLO partonic fragmentation asymmetries $A_f\equiv\Delta 
D^{\Lambda}_f/D^{\Lambda}_f$ for $f=s$, $u=d$, and $g$ at $Q^2=10\,
{\mathrm{GeV}}^2$. In the LO plot we also show (dot-dashed lines) 
the effect of assuming a maximally polarized gluon distribution at the 
initial scale for scenario 1. The NLO results in this case are very similar 
to the LO ones and are therefore not shown.
\item[Fig.\ 6] a) Compilation of the original SIDIS data \cite{h1dis} in 
terms of $x_F$. b) Comparison of our LO predictions with the data converted to
the variable $z$ (see text).
\item[Fig.\ 7] LO and NLO predictions for the SIDIS asymmetry for 
unpolarized protons and polarized $\Lambda$'s and leptons 
(see text) for our three distinct scenarios of polarized fragmentation 
functions. In a) we also show the expected statistical errors for such a 
measurement at HERA, assuming a luminosity of $500\,{\mathrm{pb}}^{-1}$,
a beam polarization of $70\%$, and a $\Lambda$ detection efficiency of 0.1. 
In b) we include the expectation for scenario 1 with a maximal gluon 
polarization at the initial scale.
\item[Fig.\ 8] The same as in Fig.\ 7, but now for HERMES kinematics.
\item[Fig.\ 9] LO and NLO predictions for the SIDIS asymmetry for polarized
protons but unpolarized leptons for two different sets of polarized
parton distributions taken from \cite{grsv}. Also shown are the expected 
statistical errors for such a measurement at HERA, calculated for the 
parameters already used for Fig.\ 7.
\end{description}
\newpage
%%%%%%%%%%%%%
% FIGURES
%%%%%%%%%%%%%
%
\pagestyle{empty}
\begin{center}

\vspace*{-4.0cm}
\hspace*{-1.5cm}
\epsfig{file=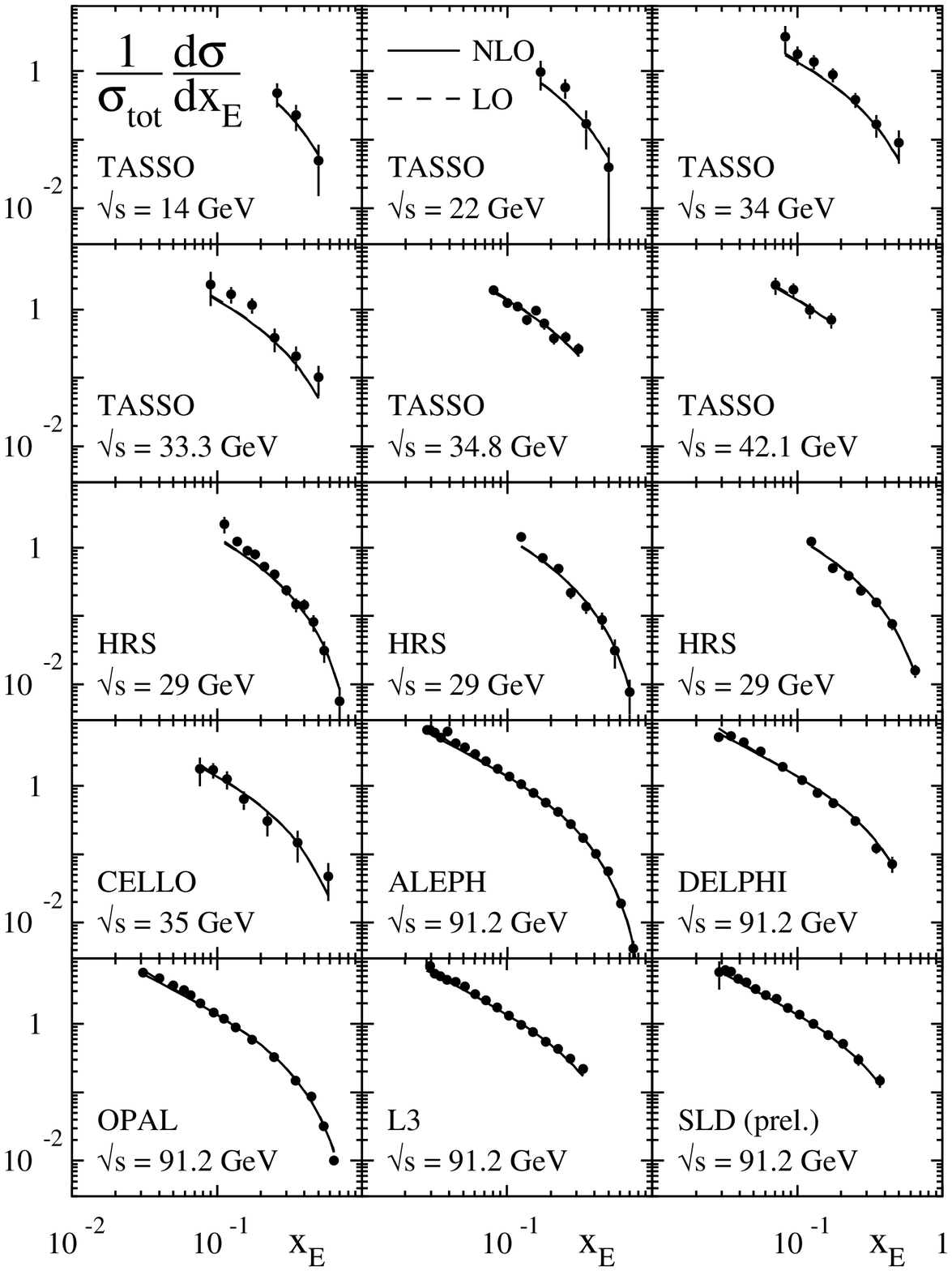}

\vspace*{-1.2cm}
\Large{\bf{Fig.\ 1}}
\end{center}

\begin{center}

\vspace*{-2.5cm}
\epsfig{file=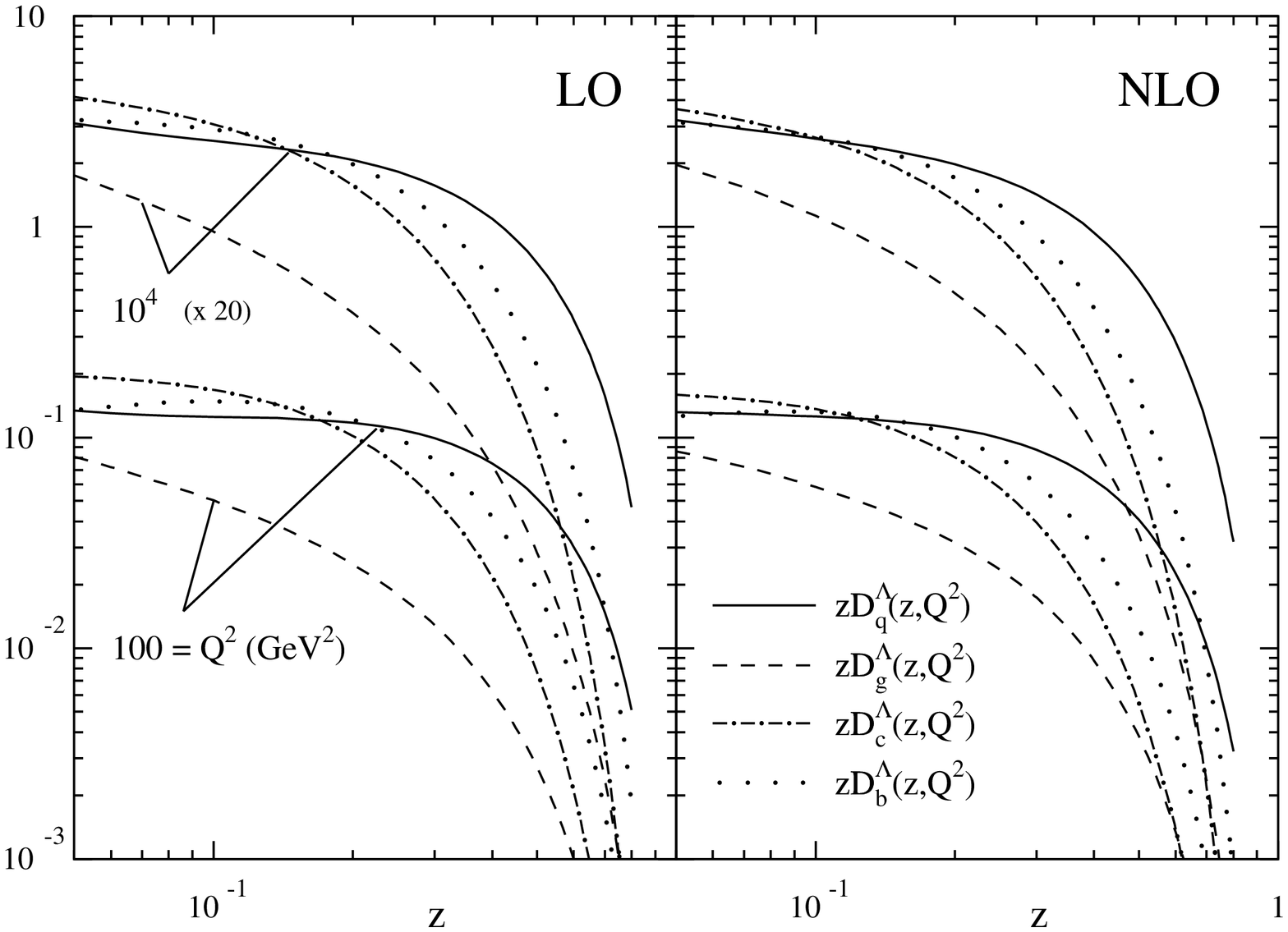,angle=90}

\vspace*{-1.2cm}
\Large{\bf{Fig.\ 2}}
\end{center}

\begin{center}

\vspace*{-2.5cm}
\epsfig{file=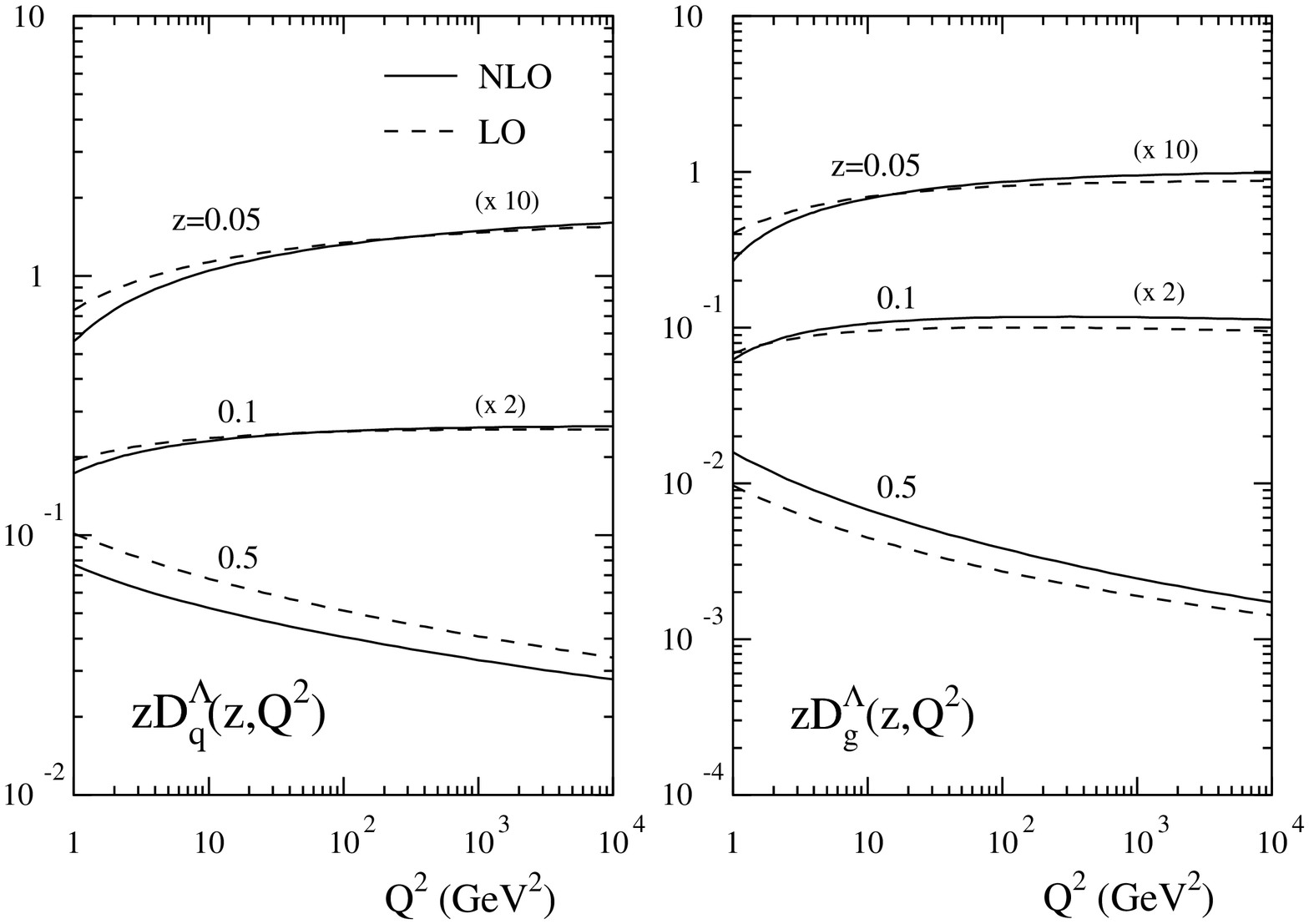, angle=90}

\vspace*{-1.2cm}
\Large{\bf{Fig.\ 3}}
\end{center}

\begin{center}

\vspace*{-2.0cm}
\hspace*{-0.8cm}
\epsfig{file=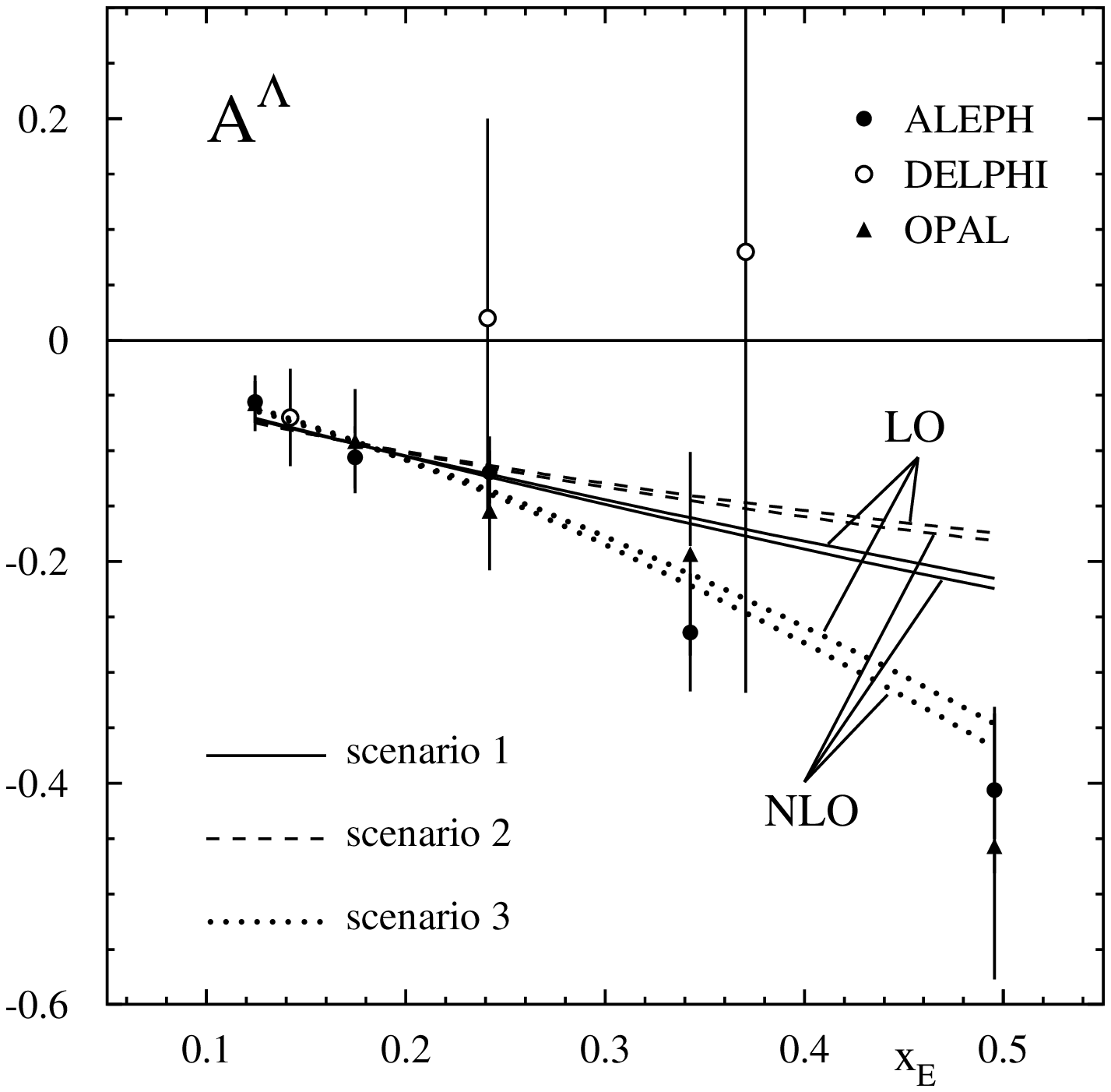}

\vspace*{-1.2cm}
\Large{\bf{Fig.\ 4}}
\end{center}

\begin{center}

\vspace*{-2.0cm}
\hspace*{-0.5cm}
\epsfig{file=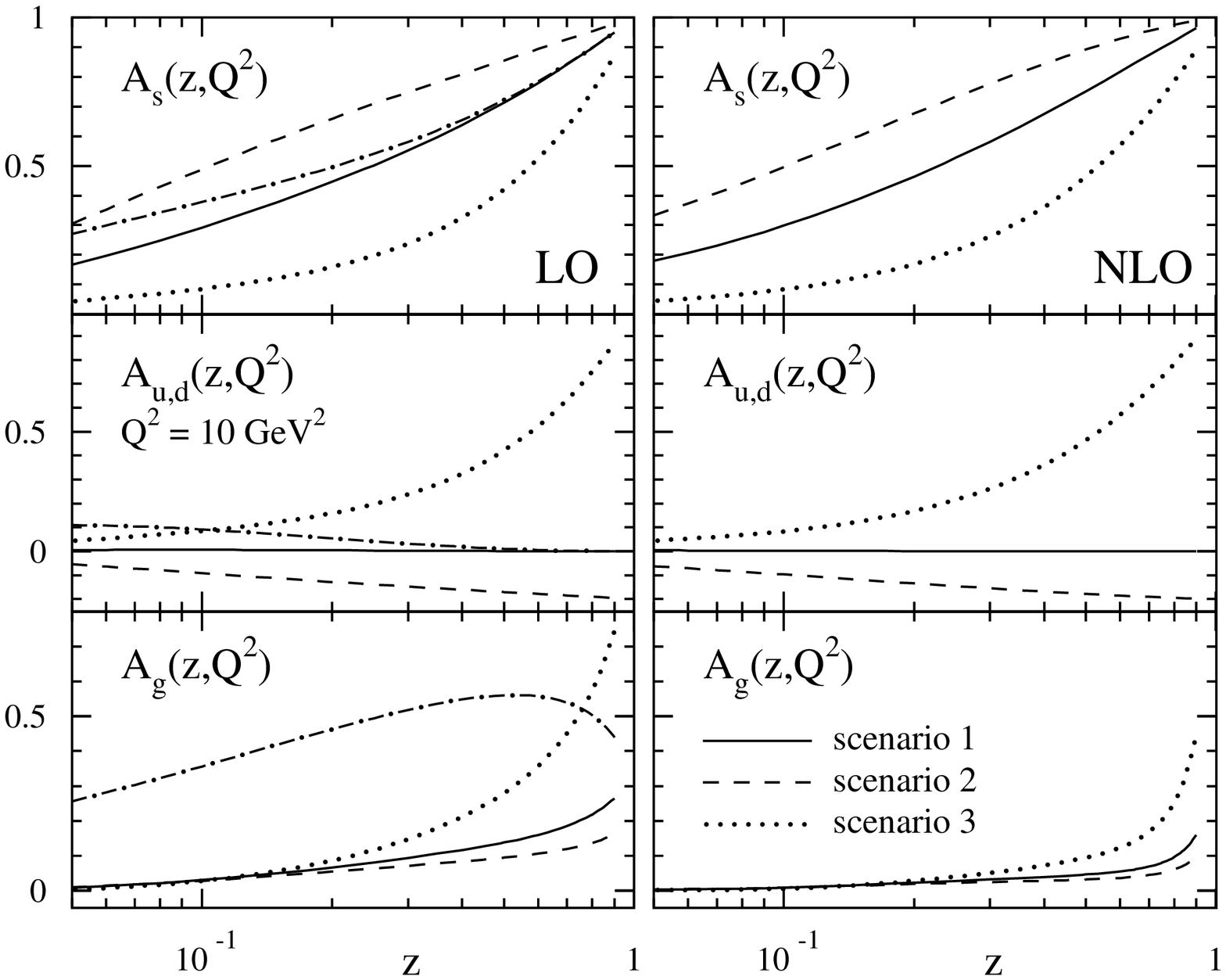, angle=90}

\vspace*{-1.2cm}
\Large{\bf{Fig.\ 5}}
\end{center}

\begin{center}

\vspace*{-2.0cm}
\hspace*{0.4cm}
\epsfig{file=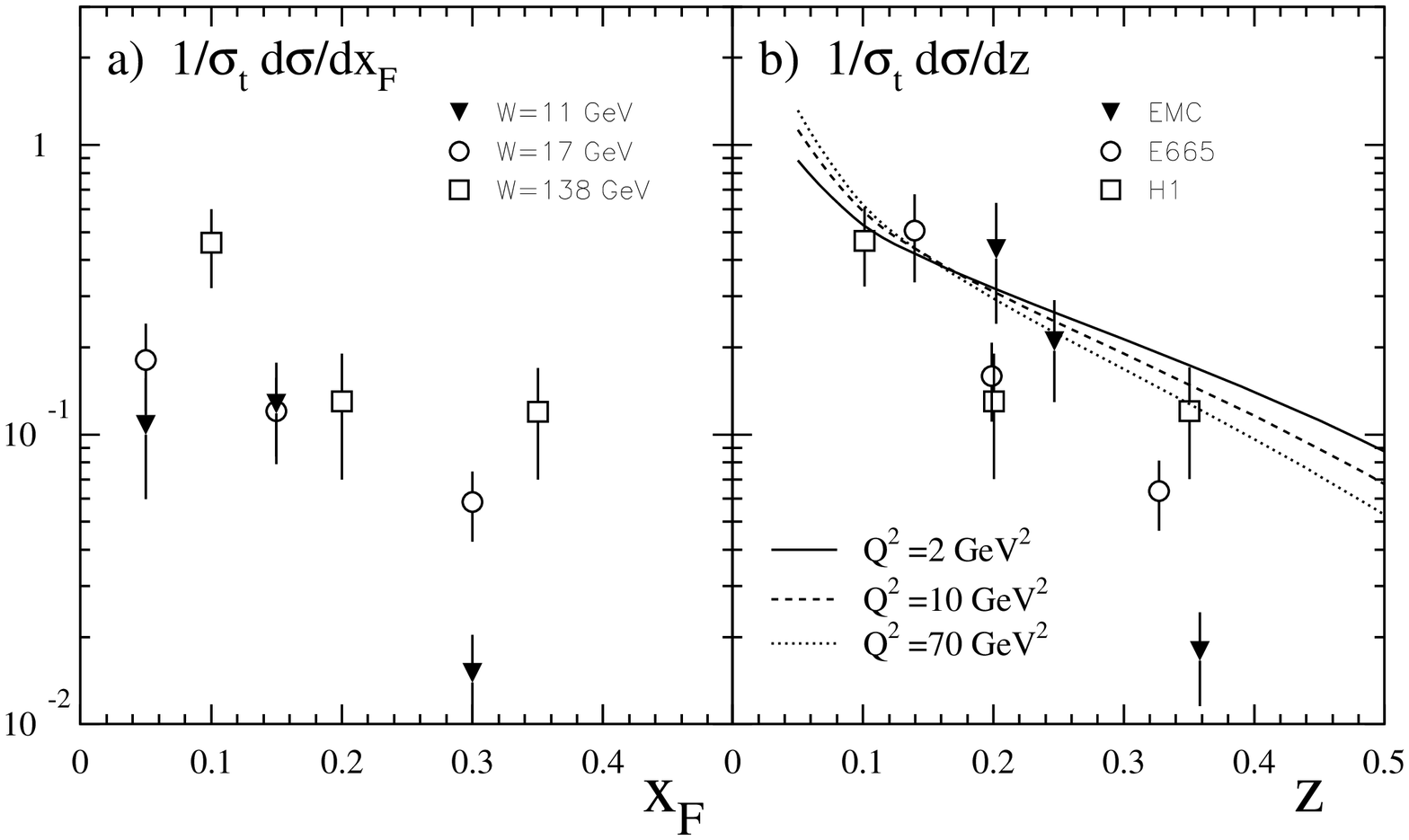, angle=90}

\vspace*{-1.2cm}
\Large{\bf{Fig.\ 6}}
\end{center}

\begin{center}

\vspace*{-2.0cm}
\hspace*{-0.9cm}
\epsfig{file=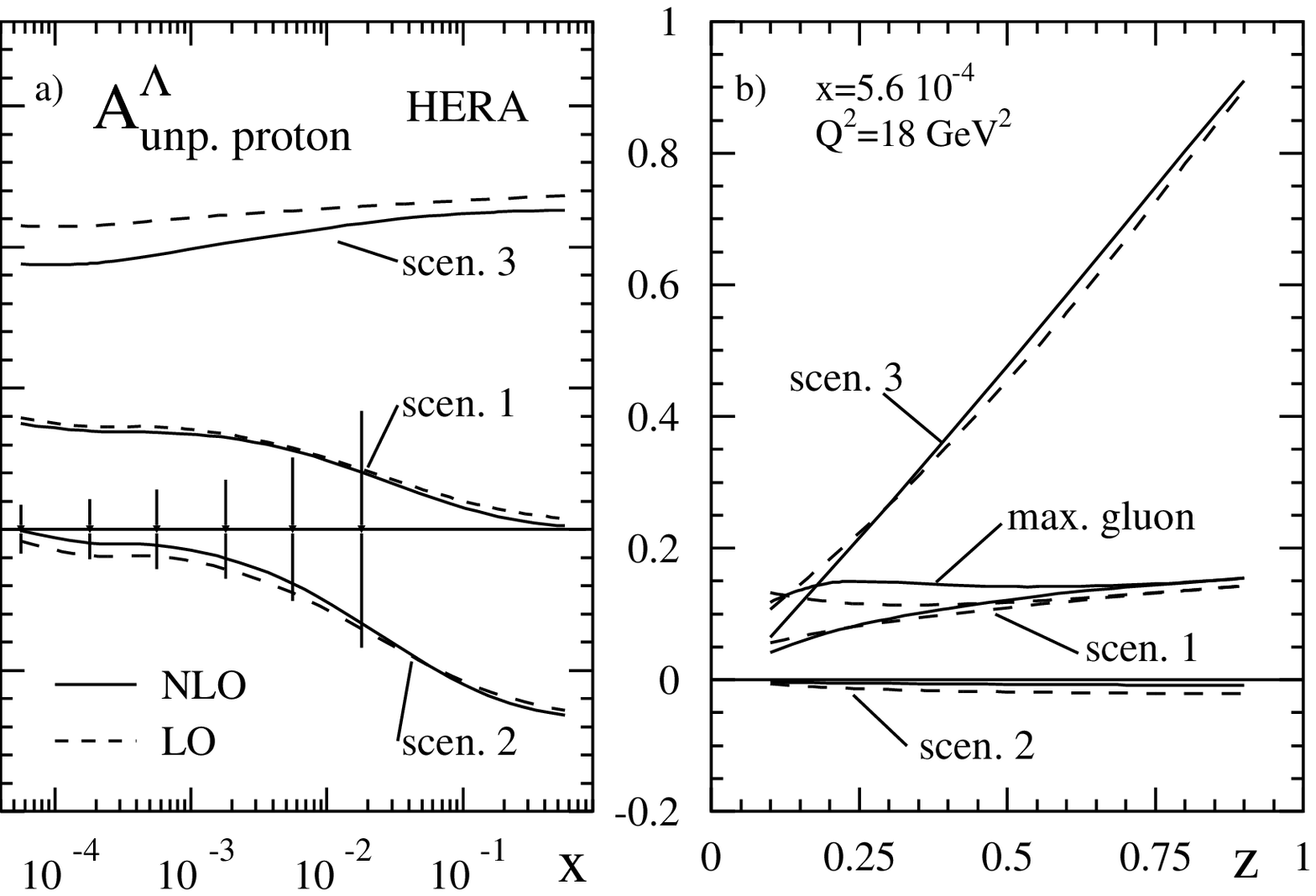, angle=90}

\vspace*{1.2 cm}
\Large{\bf{Fig.\ 7}}
\end{center}

\begin{center}

\vspace*{-2.0cm}
\hspace*{-0.9cm}
\epsfig{file=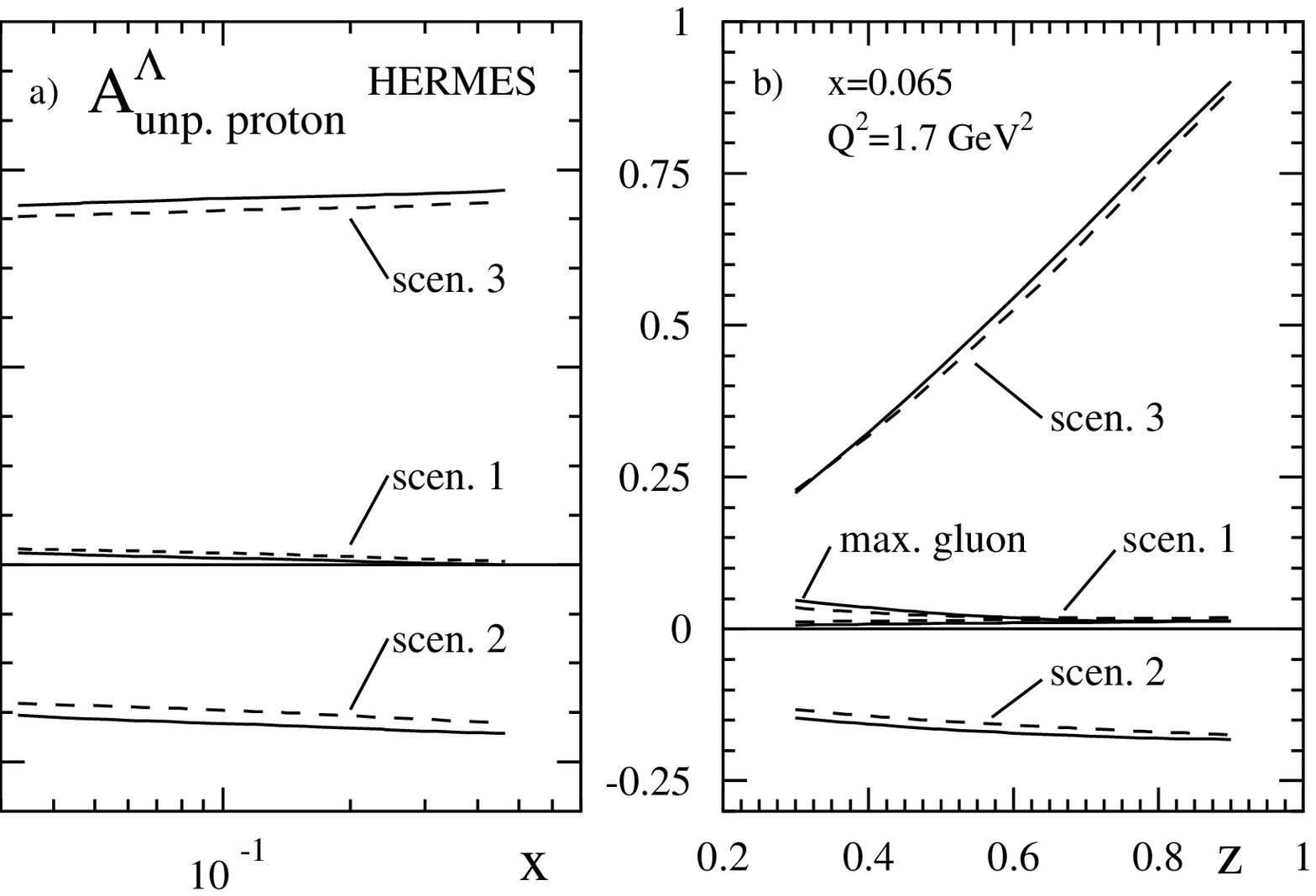, angle=90}

\vspace*{1.2cm}
\Large{\bf{Fig.\ 8}}
\end{center}

\begin{center}

\vspace*{-2.0cm}
\hspace*{-0.3cm}
\epsfig{file=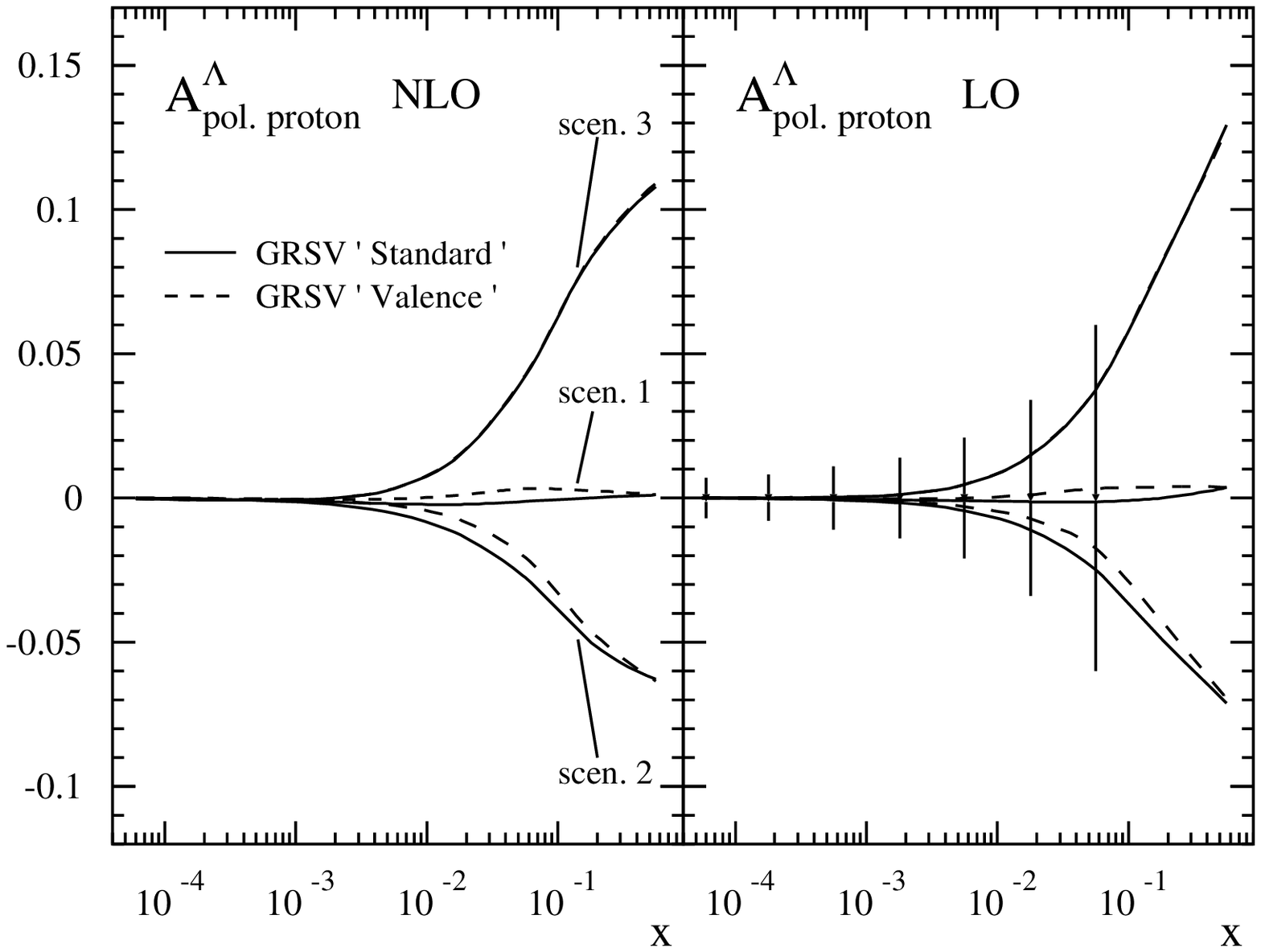, angle=90}

\vspace*{0.3cm}
\Large{\bf{Fig.\ 9}}
\end{center}

\end{document}